\documentclass[a4paper, 10pt]{article}
\usepackage[left=0.7 in, right=0.7in, top=1in, bottom = 1in]{geometry}

\usepackage{amsmath}
\usepackage{amssymb}
\usepackage{bbm}
\usepackage{graphicx}
\usepackage{amsthm}
\usepackage{color}
\usepackage{graphics}
\usepackage{epsfig}
\usepackage{cancel}
\usepackage{epstopdf}
\usepackage{hyperref}
\usepackage{xcolor}
\usepackage{cleveref}
\usepackage{authblk}
\usepackage{mathrsfs}
\usepackage{setspace}
\usepackage{moresize}
\usepackage{rotating}
\hypersetup{
colorlinks,
linkcolor={blue!100!black},
citecolor={blue!80!black},
urlcolor={blue!80!black}
}
\usepackage{natbib}

\newcommand{\mbf}[1]{\mathbf{#1}}

\newcommand{\mcal}[1]{\mathcal{#1}}

\DeclareMathOperator{\lm}{\lambda}
\DeclareMathOperator{\dl}{\Delta}

\usepackage{array}
\newcolumntype{L}[1]{>{\raggedright\let\newline\\\arraybackslash\hspace{0pt}}m{#1}}
\newcolumntype{C}[1]{>{\centering\let\newline\\\arraybackslash\hspace{0pt}}m{#1}}
\newcolumntype{R}[1]{>{\raggedleft\let\newline\\\arraybackslash\hspace{0pt}}m{#1}}

\newcommand{\tc}[1]{\textcolor{blue}{#1}}
\title{Wrinkling as a mechanical instability in  growing annular hyperelastic plates}
\author[1]{{Sumit Mehta}}
\author[1]{{Gangadharan Raju}}
\author[2]{{Prashant Saxena}\thanks{Corresponding author email: prashant.saxena@glasgow.ac.uk}}

\affil[1]{Department of Mechanical and Aerospace Engineering \protect\\
Indian Institute of Technology Hyderabad, India}
\affil[2]{James Watt School of Engineering,  University of Glasgow, Glasgow G12 8LT, UK}
\date{}

\allowdisplaybreaks  

\numberwithin{equation}{section}
\begin{document}
\maketitle
\begin{abstract}
Growth-induced instabilities are ubiquitous in 
biological systems and lead to diverse morphologies in the form of wrinkling, folding, and creasing. The current work focusses on the mechanics behind growth-induced wrinkling instabilities in an incompressible annular hyperelastic plate. 
The governing differential equations for a two-dimensional plate system are derived using a variational principle with no apriori kinematic assumptions in the thickness direction.
A linear bifurcation analysis is performed to investigate the stability behaviour of the growing hyperelastic annular plate by considering both axisymmetric and asymmetric perturbations.
The resulting differential equations are then solved numerically using the compound matrix method to evaluate the critical growth factor that leads to wrinkling.
The effect of boundary constraints, thickness, and radius ratio of the annular plate on the critical growth factor is studied.
For most of the considered cases, an asymmetric bifurcation is the preferred mode of instability for an annular plate.
Our results are useful to model the physics of wrinkling phenomena in growing planar soft tissues, swelling hydrogels, and pattern transition in two-dimensional films growing on an elastic substrate.
\end{abstract}
\vspace{1in}

\section{Introduction}
The growth process typically involves a change of body mass and generation of residual stresses in an evolving system which induces large deformation that triggers the mechanical instabilities. These growth-induced instabilities 
lead to the formation of diverse patterns in the form of wrinkling, folding, creasing \citep{li2012mechanics} that are critical to biological systems such as plants \citep{dai2014critical, coen2004genetics}, 
tissues, and organs \citep{ambrosi2011perspectives}.
Moreover, certain anomalies {in human biological systems} like narrowing airways due to asthma \citep{wiggs1997mechanism}, and malformation of the cerebral cortex \citep{raybaud2011development} give rise to patterns that defines a new physiological function  of the biological system. 
In addition, to avoid scar formation during surgeries, irregular wrinkles generated during cutaneous wound healing are studied \citep{cerda2005mechanics, nassar2012calpain}.
Besides biological applications, pattern formation and their transition during growth and remodelling have been extensively studied from a purely mechanics aspect of view
\citep{amar2005growth, liang2011growth, cao2012biomechanical, budday2014role, balbi2013morpho, limbert2018skin}.
Instabilities in soft solids such as swollen hydrogels \citep{ionov2013biomimetic} and elastomers \citep{cao2012wrinkles, kempaiah2014nature} can be tailored to create desired patterns and have found applications in the design of stretchable electronics \citep{khang2009mechanical}, smart morphable surfaces in aerodynamics control \citep{terwagne2014smart}, wearable communication devices \citep{rogers2010materials}, and shape-shifting structures \citep{stein2019buckling}.
Therefore, it is fundamentally important to understand the mechanics of growth that regulate the instabilities in soft growing bodies. 
To accomplish this, a consistent mathematical model based on the continuum mechanics framework is used for studying growth-induced instabilities in morpho-elastic structures \citep{ kuhl2014growing, goriely2017mathematics}.

To describe the kinematics of growth, the total deformation gradient is decomposed into a growth tensor and an elastic deformation tensor \citep{rodriguez1994stress} ignoring the initial residual stress configuration \citep{du2018modified, du2019influence}. 
The former tensor describes the local change in volume with the addition/subtraction of material as well as the deformation incompatibility due to non-uniform growth in the neighbourhood of a material point \citep{garikipati2004continuum, goriely2007definition}. 
The elastic deformation tensor is required to ensure compatibility and this generally results in residual stresses that can trigger mechanical instabilities.
Based on this principle, growth-induced instabilities have been extensively studied in soft biological tissues  \citep{amar2005growth, goriely2005differential, moulton2011circumferential, li2012mechanics, wu2015growth, liu2020surface, liu2021influence} as well as swelling hydrogels \citep{li2013tissue}.

Numerous works employ a hyperelastic membrane model to study instabilities in thin soft tissues \citep{papastavrou2013mechanics,swain2015interfacial, swain2016mechanics}. 
\cite{jia2018curvature} studied the bifurcation behaviour of thin growing film on a cylindrical substrate. They showed that curvature delays the bifurcation point and requires a high magnitude of growth factor to induce the circumferential wrinkling instability on a curved surface.
Recently, \cite{wang2020wrinkling} developed a model to describe the nonlinear behaviour of hyperelastic curved shells under finite strain. They explored the effect of curvature on the tunability of wrinkling and smoothing regimes, and the post-buckling evolution of a highly stretched soft shell.
Above works show the interplay between growth and elasticity induces large deformation and one needs to adopt consistent plate or shell theories to capture the combined effect of bending and stretching deformation. 
Classical plate theories like Kirchhoff-Love, F\"oppl-von K\'arm\'an (FvK), and Mindlin-Reisner theory have been widely used to investigate the bifurcation behaviour of thin elastic structures \citep{coman2006localized, coman2015asymptotic, li2010buckling}, and liquid crystal elastomers \citep{mihai2020plate}.
\cite{dervaux2009morphogenesis} developed a FvK plate theory
to investigate large deformation and growth-induced patterns in a thin hyperelastic plate.
\cite{efrati2009elastic} proposed the geometric theory for non-Euclidean thin plates to capture the large displacements in thin growing bodies where growth deformations are interpreted as the  evolution metric. 
Based on this theory, \cite{pezzulla2016geometry} studied the mechanics of thin growing bilayers by describing the geometry of mid surface with first and second fundamental form and \cite{dias2011programmed} presented the inverse approach to investigate the growth (swelling) patterns of initially prescribed axisymmetric shapes. \cite{jones2015optimal} developed the numerical framework to determine the optimal distribution of growth stresses by specifying the deformed target shape of the plate using FvK theory. \cite{holmes2019elasticity} performed post-bifurcation analysis to investigate the instability phenomena in thin soft materials. Further, \cite{mora2006buckling} experimentally and theoretically investigated the patterns arising from differential swelling of gels to mimic the behaviour of growth-induced deformation in biological tissues.
However, these theories are based on apriori assumptions of displacement variation along the thickness of the plate and therefore have certain limitations. This has led to the development of a reduced 2-D plate theories with no apriori kinematic assumptions that are consistent with the 3-D elasticity theory.

\cite{kienzler2002consistent} proposed a consistent asymptotic plate theory based on linear elasticity which does not apply any kinematic assumptions and all the unknowns are treated as independent variables.
\cite{dai2014consistent} proposed a finite strain plate theory for compressible hyperelastic materials based on the principle of minimisation of potential energy which can achieve term-wise consistency without any ad-hoc assumption for general loading conditions.
\cite{wang2016consistent} extended this approach to incompressible hyperelastic materials with an additional Lagrange multiplier to accommodate the incompressibility constraint.
\cite{wang2018consistent} derived a consistent finite-strain plate theory for growth-induced large deformation and studied the buckling and post-buckling behaviour of a thin rectangular hyperelastic plate subjected to axial growth.
They also showed that the consistent theory reduces to the FvK theory in the limit of small plate thickness.
Recently, the finite strain asymptotic plate theory has been applied to study the plane strain problems of growth-induced deformation in single and multi-layered hyperelastic plates \citep{wang2019shape, du2020analytical}. 
\cite{wang2022theoretical} studied the inverse approach of determining the inhomogeneous growth fields of target 3-D shapes by using stress-free \citep{chen2020stress} finite strain theory for thin hyperelastic plates.
\cite{liu2020consistent} discussed the large elastic deformation in nematic liquid crystal elastomer and the authors of the present paper \citep{mehta2021growth} investigated the bifurcation behaviour of circular hyperelastic plate under the influence of growth using finite strain asymptotic plate theory. 

Instability behaviour of highly deformable soft tissues is best understood by a bifurcation analysis of the corresponding system of partial differential equations (PDEs) \citep{vandiver2009differential}.
Generally an analytical solution is not possible and numerical approaches based on the finite element method are appropriate \citep{saez2016theories, dortdivanlioglu2017computational}.
Specialised finite element approaches have been employed to compute growth-induced deformation in soft materials while avoiding the volumetric locking arising from the assumption of incompressibility. 
\cite{zheng2019solid} developed the solid-shell based finite element model to investigate the growth deformation in incompressible thin-walled soft structures. \cite{groh2022morphoelastic} developed the seven parameter quadrilateral shell element to avoid locking phenomena in shells and investigated the growth-induced instability in slender structures. 
\cite{kadapa2021advantages} proposed the mixed displacement-pressure finite element formulation to study the compressible and incompressible deformation in growth problems. 
A virtue of the plate theory used in this paper, on the other hand, is that it works well with both compressible and incompressible materials without imposing any kinematic assumptions.
We provide quantitative comparisons obtained via our approach with existing computational results in the literature.
\begin{figure}
\centering
\includegraphics[width=0.5\linewidth]{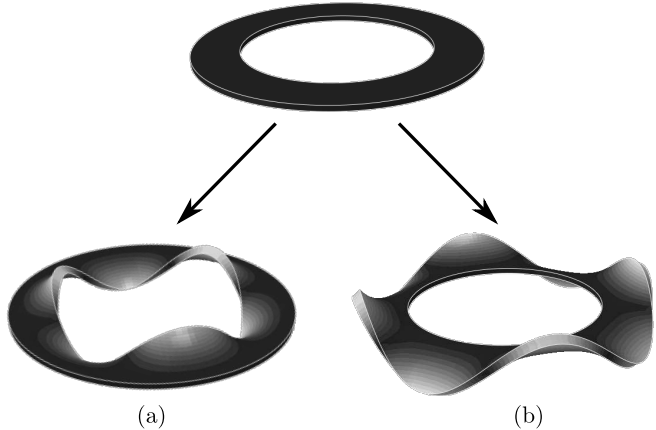}
\caption{Possible out-of-plane wrinkled configurations of a hyperelastic annular plate growing with in-plane growth function when subjected to different type of boundary conditions i.e., (a) constrained outer boundary and unconstrained inner boundary, and (b) constrained inner boundary and unconstrained outer boundary. The reference configuration (left) is a flat plate and the deformed configurations (right) are the buckled/wrinkled configuration. In each case, the unconstrained boundary exhibits wrinkles induced by growth.} \label{buckling_of_plate}
\end{figure}

The current work investigates wrinkling phenomena in growing hyperelastic annular plates that are appropriate models for human tissues and plants \citep{liang2009shape, steele2000shell} when subjected to different boundary constraints as shown in Figure \ref{buckling_of_plate}. 
For example, (a) wrinkling patterns around contracting skin wound subjected to constrained outer edge conditions \citep[Figure 5]{flynn2008simplified} (b) wrinkle formation in a circular shaped leaf when the stiff inner part can be approximated by a clamped boundary condition.
Note that we limit our discussion to only out of plane wrinkling instabilities and have not considered local instabilities like folding \citep{tallinen2015mechanics} or creasing \citep{jin2011creases, wang2015three, yang2021perturbation}.
In general,  growing systems are more appropriately modelled by anisotropic and inhomogeneous (differential) growth laws that lead to diverse patterns \citep{ ambrosi2011perspectives, huang2018differential} governing the final shape of the system. 
However, in this work,  we assume an externally-driven homogeneous isotropic growth law to focus on the mechanics of growth in annular plate structures.
The growth function is considered as a control parameter responsible for the change of shape \citep{wang2019shape, li2022analytical} and the onset of instability. 
We also choose a neo-Hookean material model that allows us to model nonlinear deformation while still keeping the resulting equations relatively simple and retaining the key aspects of the mechanics of the system.
We have applied the consistent finite strain plate theory introduced by \cite{wang2018consistent} to derive the governing differential equations (GDEs) for general loading conditions.
Subsequently, we perform the stability analysis  by perturbing the principal solution 
subjected to two different cases of boundary conditions. 
In the first case, the inner boundary of the annular plate is unconstrained and the outer boundary is constrained. 
For the second case, the inner boundary of the plate is constrained and the outer boundary is unconstrained. 
The resulting nonlinear ordinary differential equations (ODEs) 
are solved numerically to evaluate the critical value of growth parameter. 
For each case of boundary conditions,
we investigate the type of perturbation namely, axisymmetric or asymmetric 
corresponding to stable bifurcation solution of the annular plate and also study the effect of plate thickness, radius ratio on the onset of instability.
  
\subsection{Organisation of this manuscript}
The remainder of this paper is organised as follows. 
In Section \ref{sec:3-D_form}, a general formulation for a three-dimensional annular plate is established and then the same is reduced to a two-dimensional system by eliminating the dependence on the thickness variable using a series approximation. 
In Section \ref{Base_sol}, we discuss the principal solution associated with the growth-induced deformation in an incompressible neo-Hookean annular plate subjected to two different boundary conditions. We validate our 2-D plate framework by comparing the obtained numerical pre-buckling solution with the existing analytical solution of a circular ring later in this section.
In Section \ref{sec:Bifurcation_analysis}, we derive the non-dimensional nonlinear ODEs associated with asymmetric as well as axisymmetric perturbations. 
Section \ref{sec:Result_sec} begins with a comparison of the current plate theory to FvK plate theory. Then, we apply the compound matrix method to solve the bifurcation problem and compare the bifurcation solutions associated with each type of perturbation and boundary conditions. At the end of this section, we provide the comparison of bifurcation solution obtained with this current plate theory with the existing solution obtained using finite element approach.
Finally, we present our conclusion in Section \ref{conclusion}. Supplementary mathematical derivations are provided in the Appendix.
\subsection{Notation}
\underline{Brackets}: Two types of brackets are used. Round brackets ( ) are used to define the functions applied on parameters or variables. Square brackets [ ] are used to clarify the order of operations in an algebraic expression. 
Square brackets are also used for matrices and tensors. At some places we use the square bracket to define the functional.

\noindent \underline{Symbols}: A variable typeset in a normal weight font represents a scalar. A lower-case bold weight font denotes a
vector and bold weight upper-case denotes tensor or matrices. 
Tensor product of two vectors $\mbf{a}$ and $\mbf{b}$ is defined as $[\mbf{a} \otimes \mbf{b}]_{ij} = [\mbf{a}]_i [\mbf{b}]_j$.
Tensor product of two second order tensors $\mbf{A}$ and $\mbf{B}$ is defined as either $[\mbf{A} \otimes \mbf{B}]_{ijkl} = [\mbf{A}]_{ij} [\mbf{B}]_{kl} $ or $[\mbf{A} \boxtimes \mbf{B}]_{ijkl} = [\mbf{A}]_{ik} [\mbf{B}]_{jl} $. 
{Higher order tensors are written in bold calligraphic font with a superscript as $ \pmb{\mathcal{A}}^{(i)}$, where superscript `$i$' tells that the function is differentiated $i+1$ times. For example, $\displaystyle \pmb{\mathcal{A}}^{(1)} = \frac{\partial f(\mathbf{A})}{ \partial \mathbf{A} \partial \mathbf{A}} $ is a fourth order tensor. Operation of a fourth order tensor on a second order tensor is denoted as $ [\pmb{\mathcal{A}}^{(1)} : \mbf{A}]_{ij} = [\pmb{\mathcal{A}}^{(1)}]_{ijkl} [\mathbf{A}]_{kl}$ }.
Inner product is defined as $\mbf{a} \cdot \mbf{b} = [\mbf{a}]_i [\mbf{b}]_i$ and $\mbf{A} : \mbf{B} = [\mbf{A}]_{ij} [\mbf{B}]_{ij}$.
The symbol $\nabla$ denotes the two-dimensional
differentiation operator.
We use the word `Div' to denote divergence in three dimensions.

\noindent \underline{Functions}: $\det(\mathbf{A})$ denote the determinant of a tensor $\mathbf{A}$. $\text{tr}(\mathbf{A})$ denote the trace of tensor $\mathbf{A}$. $\text{diag}(a, b, c)$ denotes a second order tensor with only diagonal entries $a, b~ \text{and}~ c$. 

\section{Governing equation with variational principle} \label{sec:3-D_form}
\begin{figure}
\centering
\includegraphics[width=0.9\linewidth]{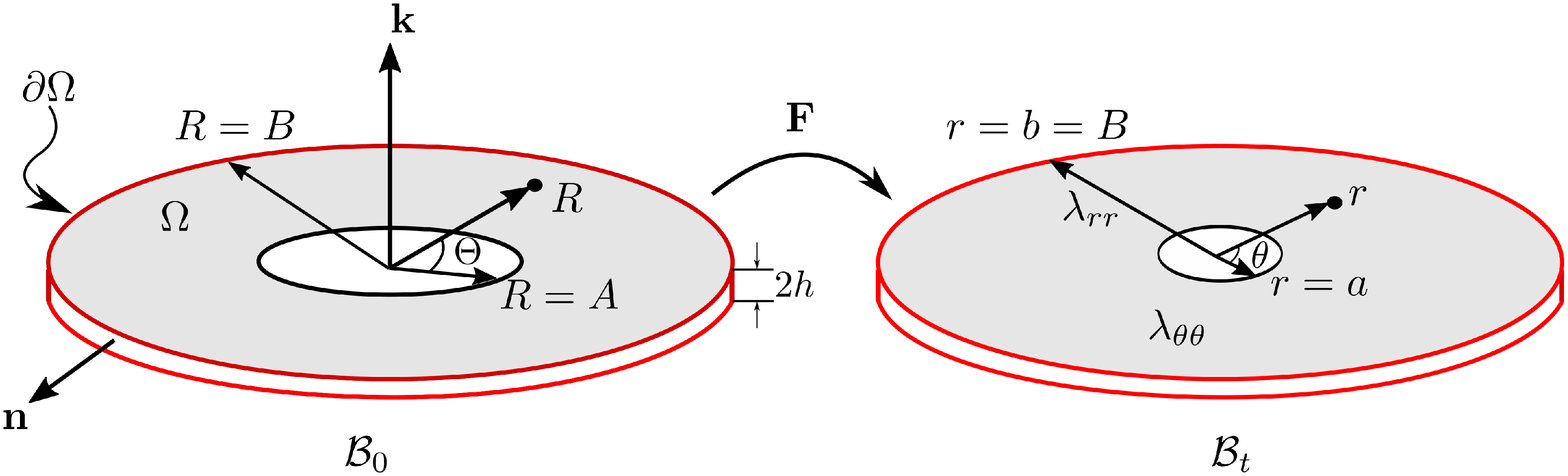}
\caption{A schematic of finite deformation of an annular plate due to growing region $\Omega$ with constrained outer boundary i.e., the outer radius is same in the reference, $\mathcal{B}_0$ (left) and in the deformed, $\mathcal{B}_t$ (right) configuration. The region $\Omega$ is growing with a factor of $\lm_{rr}$ and $\lm_{\theta \theta}$ in the radial and hoop directions, respectively.}
\label{Annulus_plate_trac}
\end{figure}

Consider a thin annular plate with constant thickness ($2h$) occupying the region $ \Omega \times [0,2h]$ in the reference configuration $\mathcal{B}_0 \in \mathscr{R}^3$ which then deforms  to the current configuration $\mcal{B}_t \in \mathscr{R}^3$ as shown in Figure \ref{Annulus_plate_trac}. Coordinates of a point in the undeformed configuration are given by $R,\Theta,Z$ and in the deformed configuration by $r, \theta, z$. 
Position vector in $\mathcal{B}_0$ is denoted as $\mathbf{X} (R, \Theta, Z) $ and denoted as $\mathbf{x} (r, \theta, z)$ in $\mcal{B}_t$.
The inner and outer radii in the reference configuration are denoted as $A$ and $B$, and the same in deformed configuration are denoted as $a$ and $b$, respectively.
Using the position vector $\mbf{X}$ and $\mbf{x}$, the deformation gradient is expressed as $
\displaystyle \mathbf{F}=\frac{\partial \mathbf{x}}{\partial {\zeta}} + \frac{\partial \mathbf{x}}{\partial Z} \otimes \mbf{k} $,
where $\zeta = R \mbf{e}_R + \Theta \mbf{e}_\Theta$ and $\mathbf{k}$ is the unit normal to the surface $\Omega$ in the reference configuration. Following the multiplicative decomposition approach proposed by \cite{rodriguez1994stress}, the total deformation gradient of a growing plate is decomposed  as $\mbf{F} = \mbf{AG}$ in $A \leq R\leq B$, where $\mathbf{G}$ represents the growth tensor and $\mathbf{A}$ represents the elastic deformation tensor.
We also assume the  material to be incompressible and therefore to follow the constraint %
$L({\mathbf{F,G}})=L_0(\mathbf{FG}^{-1}) = \det(\mathbf{A})-1=0$.
The energy density ($\phi$) per unit volume of the  material is assumed as $\phi (\mbf{F, G}) = J_G \phi_0(\mbf{FG}^{-1})$,
where $J_G = \det(\mathbf{G})=\det(\mathbf{F})$ describes the local change in volume due to growth and $\phi_0 (\mathbf{FG}^{-1})$ is the elastic strain energy density.
The internal potential energy functional ($\psi$) for the incompressible plate is 
\begin{align}
\psi[\mathbf{x}(\mathbf{X}), p(\mathbf{X})] = \int_{\Omega} \int_{0}^{2h} J_G \phi_0 (\mathbf{FG}^{-1}) dV - \int_{\Omega} \int_{0}^{2h} \big[ J_G~ p(\mathbf{X}) L_0(\mathbf{FG}^{-1})\big]dV ,
\end{align}
where $p(\mathbf{X})$ is the Lagrange multiplier associated with the incompressibility constraint.
We apply the principle of minimum potential energy and
vanishing of the first variation with respect to $\mathbf{x}$ and $p$ of the above functional results in the GDEs with traction free boundary conditions
\begin{subequations}
\begin{align}
& \text{Div}~\mathbf{P}=\mathbf{0}, \qquad \qquad \qquad \qquad \qquad \qquad \qquad  \qquad \qquad ~~ \text{in} ~\Omega \times  [0,2h], \label{equlibrium_eq1}\\
&\mathbf{Pk}\big|_{Z=0} = \mbf{0}, \qquad \left.\mathbf{Pk} \right|_{Z=2h} = \mbf{0}, \qquad \qquad \qquad \qquad  \qquad  ~ \text{on}~\Omega, \label{equlibrium_eq2} \\
& \left. \mathbf{P n} \right|_{R = A} = \mbf{0}, \qquad \left. \mathbf{P n} \right|_{R = B} = \mbf{0}, \qquad \qquad   \qquad \qquad \qquad  \text{on}~ \partial  \Gamma \times [0,2h], \label{equlibrium_eq3}
\end{align}
\end{subequations}
and the incompressibility constraint, i.e., $\det(\mbf{A}) - 1 = 0$. Here, $\mathbf{n}$ is the unit outward normal to the lateral boundary ($\partial \Gamma$) and $\mathbf{P} = J_G \left[\displaystyle\frac{\partial \phi_0}{\partial \mathbf{A}}-p \frac{\partial L_0}{\partial \mathbf{A}}\right]\mathbf{G}^{-T}$ in $\Omega$ is recognised as the first Piola Kirchhoff stress tensor.

\subsection{Two-Dimensional plate model}
To obtain the 2-D formulation for the annular plate, we apply a series expansion of the unknown variables, $\mbf{x}$ and $p$ about the bottom surface of plate ($Z = 0$) along the thickness direction following the approach by \cite{wang2018consistent, wang2019shape}
\begin{align}
\mbf{x}(\mbf{X}) = \sum\limits_{n=0}^{3} \frac{Z^n}{n!} \mbf{x}^{(n)}(\zeta) + O(Z^4), \quad \text{and} \quad p(\mbf{X}) = \sum\limits_{n=0}^{3} \frac{Z^n}{n!} p^{(n)}(\zeta) + O(Z^4), \label{def_cord_exp}
\end{align}
where we have used the notation $\mbf{x}^{(n)} = \displaystyle\frac{\partial^n \mbf{x}}{\partial Z^n}$ and $p^{(n)} = \displaystyle\frac{\partial^n p}{\partial Z^n}$.
Using \eqref{def_cord_exp}, we obtain a recursion relation for the deformation gradient as, $\mbf{F}^{(n)} = \nabla \mbf{x}^{(n)} + \mbf{x}^{(n+1)} \otimes \mbf{k} $. 
Similarly, we expand the elastic tensor $\mbf{A}$, and inverse transpose of growth tensor ($\mbf{G}^{-T}$) to obtain the Piola Kirchhoff stress tensor $\mbf{P}$  (see \eqref{app:series_exp}). 
Upon neglecting the body force and external traction, the equilibrium equation \eqref{equlibrium_eq1} is given as  the recursive relation 
\begin{align}
\nabla\cdot {\mathbf{P}}^{(n)}+ {\mathbf{P}}^{(n+1)}\mathbf{k}=\mathbf{0}, \qquad \text{for} \qquad n = 0,~1,~2. \label{recurssion_stress}
\end{align}
Substituting $\mbf{A} = \mbf{FG}^{-1}$ in the expression of Piola stress obtained as $\mathbf{P} = J_G \left[\displaystyle\frac{\partial \phi_0}{\partial \mathbf{A}}-p \frac{\partial L_0}{\partial \mathbf{A}}\right]\mathbf{G}^{-T}$ and making use of \eqref{recurssion_stress}, we obtain the explicit expressions of $\mbf{P}^{(0)}$, $\mbf{P}^{(1)}$, and $\mbf{P}^{(2)}$
which in component form ($[\mathbf{P}]_{ij} = P_{ij}$) are given as
\begin{subequations} \label{comp_Piola}
\begin{align}
P_{ij}^{(0)}& = J_G\bigg[\pmb{\mathcal{A}}^{(0)}-p^{(0)}\pmb{\mathcal{L}}^{(0)}\bigg]_{i\alpha}\bar{G}^{(0)}_{\alpha j}, \label{Piola_0}\\
P_{ij}^{(1)}&=J_G \left[\bigg[\left[\pmb{\mathcal{A}}^{(1)}-p^{(0)} \pmb{\mathcal{L}}^{(1)}\right]_{i \alpha k \beta} A^{(1)}_{k \beta }-p^{(1)} \mathcal{L}^{(0)}_{i \alpha}\bigg]\bar{G}^{(0)}_{\alpha j} + \left[\pmb{\mathcal{A}}^{(0)}-p^{(0)} \pmb{\mathcal{L}}^{(0)}\right]_{i\alpha}\bar{G}^{(1)}_{\alpha j}\right],\label{Piola_1}\\
P^{(2)}_{ij}&=J_{G}\bigg[\bigg[{\mathcal{A}}^{(1)}_{i k \alpha \beta}A^{(2)}_{\alpha \beta}+\big[\pmb{\mathcal{A}}^{(2)}-p^{(0)} \pmb{\mathcal{L}}^{(2)}\big]_{i k \alpha \beta m n}A^{(1)}_{\alpha \beta}A^{(1)}_{m n}-2p^{(1)}\mathcal{L}^{(1)}_{i k \alpha \beta}A^{(1)}_{\alpha \beta}-p^{(0)}\mathcal{L}^{(1)}_{i k \alpha \beta}A^{(2)}_{\alpha \beta}\nonumber \\
  & -p^{(2)}\mathcal{L}^{(0)}_{ik}\bigg]\bar{G}^{(0)}_{kj}+\bigg[2\mathcal{A}^{(1)}_{i k \alpha \beta}A^{(1)}_{\alpha \beta}-2 p^{(0)}\mathcal{L}^{(1)}_{i k \alpha \beta}A^{(1)}_{\alpha \beta}-2 p^{(1)}\mathcal{L}^{(0)}_{ik}\bigg] \bar{G}^{(1)}_{k j} +  \bigg[\pmb{\mathcal{A}}^{(0)}-p^{(0)}\pmb{\mathcal{L}}^{(0)}\bigg]_{i k}\bar{G}^{(2)}_{kj}\bigg],\label{Piola_2}
\end{align}
\end{subequations}
with $\displaystyle \pmb{\mathcal{A}}^{{(i)}}(\mathbf{A}^{(0)})= \left.\frac{\partial^{i+1} \phi_0(\mathbf{A})}{\partial \mathbf{A}^{i+1}}\right|_{\mathbf{A}=\mathbf{A}^{(0)}}$ and~ $\displaystyle \pmb{\mathcal{L}}^{(i)}(\mathbf{A}^{(0)})=\left.\frac{\partial L_0(\mathbf{A})}{\partial \mathbf{A}^{i+1}}\right|_{\mathbf{A}=\mathbf{A}^{(0)}}$. 
Further mathematical details of the above calculations are provided in the supplementary document. 
Equations \eqref{Piola_1} -- \eqref{Piola_2} involve higher derivatives of $\mbf{A}$ which are obtained using the series expansion of $\mbf{F}$ and $\mbf{G}^{-T}$ 
\begin{align}
\mathbf{A}^{(0)} = \mathbf{F}^{(0)}\bar{\mathbf{G}}^{{(0)}^{T}}, \quad 
\mathbf{A}^{(1)} = \mathbf{F}^{(0)}\bar{\mathbf{G}}^{{(1)}^{T}}+\mathbf{F}^{(1)} \bar{\mathbf{G}}^{{(0)}^{T}}, \quad 
\mathbf{A}^{(2)} =\mathbf{F}^{(0)}\bar{\mathbf{G}}^{{(2)}^{T}}+2\mathbf{F}^{(1)}\bar{\mathbf{G}}^{{(1)}^{T}}+\mathbf{F}^{(2)}\bar{\mathbf{G}}^{{(0)}^{T}}. \label{eq:exp_elastic_def}
\end{align} 
The stress-free boundary conditions on the bottom and top surfaces of the annular plate are obtained from \eqref{equlibrium_eq2} as
\begin{subequations} 
\begin{align}
\left.\mathbf{P} \mathbf{k} \right|_{Z=0}& = \mathbf{P}^{(0)}(\mathbf{A})\mathbf{k}=\mbf{0},\label{bottom_trac}\\
\left.\mathbf{P}\mathbf{k}\right|_{Z=2h} &= \mathbf{P}^{(0)}\mathbf{k}+2h\mathbf{P}^{(1)}\mathbf{k}+2h^2\mathbf{P}^{(2)}\mathbf{k} + O(h^3) = \mbf{0}, \label{top_trac}
\end{align}
\end{subequations}
and the conditions associated with the incompressibility constraint upto second order are given by 
\begin{equation}
\begin{aligned}
L_0(\mathbf{A}^{(0)})=0, \qquad & \pmb{\mathcal{L}}^{(0)}[\mathbf{A}^{(1)}]=0, \qquad \pmb{\mathcal{L}}^{(0)}[\mathbf{A}^{(2)}] + \pmb{\mathcal{L}}^{(1)}[\mathbf{A}^{(1)},~\mathbf{A}^{(1)}]=0,\\
 & \pmb{\mathcal{L}}^{(0)}[\mathbf{A}^{(3)}]+3 \pmb{\mathcal{L}}^{(1)}[\mathbf{A}^{(1)},~\mathbf{A}^{(2)}]+ \pmb{\mathcal{L}}^{(2)}[\mathbf{A}^{(1)},~\mathbf{A}^{(1)},~\mathbf{A}^{(1)}]=0,
\end{aligned}
\end{equation}
where $\pmb{\mathcal{L}}^{(0)}[\mathbf{A}^{(1)}]=\pmb{\mathcal{L}}^{(0)}:\mathbf{A}^{(1)}=\displaystyle \text{det}(\mbf{A}) \mbf{A}^{-T}:\mbf{A}^{(1)}$.
On subtracting the top \eqref{top_trac} and bottom \eqref{bottom_trac} traction conditions we obtain the 2-D plate GDE 
\begin{align}
\nabla \cdot \widetilde{\mbf{P}} = \mbf{0}.  \label{eqbm_plate_eq}
\end{align}
Here, $\widetilde{\mbf{P}} = \mbf{P}^{(0)} + h \mbf{P}^{(1)} + \displaystyle \frac{2}{3} h^2 \mbf{P}^{(2)}$ is the average stress obtained by simply taking the integration over the thickness of the plate, $ \displaystyle \mbf{\widetilde{P}} = \frac{1}{2h} \int_{0}^{2h} \mbf{P} dZ$. 
Using the series expansion approach, the equilibrium equation \eqref{eqbm_plate_eq} is expressed as 
\begin{equation}
\left.
\begin{aligned}
\nabla \cdot \mathbf{P}^{(0)}_{t} + h \nabla \cdot \mathbf{P}^{(1)}_{t} + \frac{2}{3} h^2 \nabla \cdot \mathbf{P}^{(2)}_{t} + O(h^3)=\mbf{0}, \\
\big[ \nabla \cdot \widetilde{\mbf{P}} \big] \cdot \mbf{k} = \nabla \cdot \left[{\mathbf{P}^{(0)}}^T \mbf{k} \right] + h \nabla \cdot \left[{\mathbf{P}^{(1)}}^T \mbf{k} \right] + \frac{2}{3} h^2 \nabla \cdot \left[{\mathbf{P}^{(2)}}^T \mbf{k} \right] + O(h^3)=0,
\end{aligned} \right\} \label{decom_plate_eq_1}
\end{equation}
where the subscript `$t$' represents the in-plane (or tangential) component of a vector or tensor.
Equation \eqref{decom_plate_eq_1} is then reduced to a refined plate equation \citep{wang2019uniformly, yu2020refined} by neglecting the contribution of $\mbf{P}^{(2)}$, but keeping terms of $O(h^2)$ that correspond to the bending energy  of the plate
\begin{subequations} \label{mod_gov_2-D_plate_eq}
\begin{align}
& \nabla \cdot \mathbf{P}^{(0)}_{t} + h\nabla \cdot \mathbf{P}^{(1)}_{t} =\mbf{0}, \label{mod_gov_in_plane_plate_eq}\\
& \nabla \cdot \left[ ({\mbf{P}^{(0)}}^{T} \mbf{k}) - ({\mbf{P}^{(0)}} \mbf{k}) \right] + h \bigg[ \nabla \cdot \left[ ({\mbf{P}^{(1)}}^{T} \mbf{k}) - ({\mbf{P}^{(1)}} \mbf{k}) \right]\bigg] + \frac{1}{3} h^2 \nabla \cdot [\nabla \cdot {\mbf{P}_t}^{(1)}] = 0,\label{mod_gov_transverse_plate_eq}
\end{align}
\end{subequations}
The explicit expressions of unknowns variables $\mathbf{x}^{(2)}$, $p^{(1)}~\text{and}~ \mathbf{x}^{(3)},~p^{(2)}$ in terms of $\mathbf{x}^{(0)},~\mathbf{x}^{(1)}~\text{and}~p^{(0)}$  
which result in a closed form system are provided in our previous work \citep{mehta2021growth}. 
One advantage of this plate theory is that if we neglect the bending term in \eqref{decom_plate_eq_1}, the plate system is reduced to a membrane system. 
By solving the annular plate system (ignoring $O(h)$ and $O(h^2)$ terms in \eqref{decom_plate_eq_1} or \eqref{mod_gov_2-D_plate_eq}) and applying an  incompressible Varga hyperelastic material model, we recover the membrane equation  obtained by \cite{swain2015interfacial}.
\section{Growth-induced deformation in  annular  plates} \label{Base_sol}
In this section, we discuss the  deformation of an annular plate undergoing isotropic growth
i.e., plate growing with equivalent constant growth factor ($\lambda$) in the radial and circumferential directions. 
The growth tensor $\mbf{G}$ then takes the form $\text{diag}(\lm_{rr},\lm_{\theta \theta},1)$ where $\lm_{rr} = \lm_{\theta \theta} = \lambda$.  
To simplify the calculation, the plate is assumed to be made up of an incompressible neo-Hookean material with elastic strain-energy function $\phi(\mbf{F},\mbf{G}) = J_G\phi_0(\mbf{A}) = J_G C_0 \big[ I_1 - 3 \big]$, where $I_1 = \text{tr}(\mbf{A}^T \mbf{A})$ and $2 C_0$ is the ground state shear modulus. 
Using \eqref{def_cord_exp}, the series approximation of unknown variables about the bottom surface in the cylindrical coordinate system is given as 
\begin{equation}
\begin{aligned}
r(R,Z) = \sum_{n=0}^{n=3} \frac{Z^n}{n!} r^{(n)}(R) + O(Z^4), \qquad \theta(R,Z) = \sum_{n=0}^{n=3} \frac{Z^n}{n!} \theta^{(n)}(R) + O(Z^4),\\ z(R,Z) = \sum_{n=0}^{n=3} \frac{Z^n}{n!} z^{(n)}(R) + O(Z^4), \qquad
p(R,Z) = \sum_{n=0}^{n=3} \frac{Z^n}{n!} p^{(n)}(R) + O(Z^4),
\end{aligned} \label{series_unknown_var}
\end{equation}
where $\displaystyle (\cdot)^{n} = \frac{\partial^n (\cdot)}{\partial Z^n}$.
The isotropic growth field with \eqref{series_unknown_var} results in 
\begin{align}
\big[\bar{\mathbf{G}}^{(0)} \big]=
\begin{bmatrix}
\displaystyle \frac{1}{\lambda} & 0 & 0\\
0 & \displaystyle \frac{1}{\lambda} & 0\\
0 & 0 & 1
\end{bmatrix},\quad
 \big[\mathbf{F}^{(0)}\big]=
\begin{bmatrix}
\displaystyle\frac{\partial r^{(0)}}{\partial R} & \displaystyle \frac{1}{R}\frac{\partial r^{(0)}}{\partial \Theta} &  r^{(1)} \vspace{5pt}\\
\displaystyle r^{(0)} \frac{\partial \theta^{(0)}}{\partial R} & \displaystyle  \frac{r^{(0)}}{R}\frac{\partial \theta^{(0)}}{\partial \Theta} & r^{(0)} \theta^{(1)} \vspace{5pt}\\ 
\displaystyle\frac{\partial z^{(0)}}{\partial R} & \displaystyle \frac{1}{R}\frac{\partial z^{(0)}}{\partial \Theta} &  z^{(1)}
\end{bmatrix},\quad
J_G=\text{det}({\mathbf{G}})=\lambda^2, \label{exp_growth_and_def_grad}
\end{align}
where $\bar{\mbf{G}}^{(0)}$ and $\mbf{F}^{(0)}$ are the first terms in the expansion of ${\mbf{G}}^{-T}$ and $\mbf{F}$, respectively.
Explicit expressions for the unknown variables can be derived as
\begin{align}
p^{(0)}=\frac{2C_0 \lambda^4}{\left|\nabla {\mathbf{x}^{(0)}}^{*}\right|^2},\quad
r^{(1)}= \displaystyle \frac{p^{(0)} \Delta x_{11}}{2 C_0 \lambda^2}, \quad
\theta^{(1)}=\frac{p^{(0)} \Delta x_{22}}{2 C_0 \lambda^2 r^{(0)}},\quad
z^{(1)}=\frac{p^{(0)} \Delta x_{33}}{2 C_0 \lambda^2}. \label{eq:exp_unkwn_var}
\end{align}
Equations associated with the derivation of \eqref{eq:exp_unkwn_var} and the terms corresponding to higher orders of $\mbf{\bar{G}}$, $\mbf{F}$ are detailed in Appendix \ref{app:series_exp_for_tensor}. 

\subsection{Pre-buckling solution } \label{sec_base_state}
The principal or pre-buckling axisymmetric solution
is given by
\begin{align}
r^{(0)}(R)=r(R), \quad \theta^{(0)} = \Theta, \quad  z^{(0)}(R) = C_Z, \label{principal_sol}
\end{align}
where $C_Z$ is a constant function. Upon substituting the principal solution \eqref{principal_sol} in the plate equation \eqref{mod_gov_2-D_plate_eq}, we obtain a fourth order equation in $r^{(0)}$. This is transformed to four first order ODEs of the form 
\begin{equation}
\mbf{D} \mbf{y}' = \mbf{q},\label{eq:sys_ODES_base} 
\end{equation} 
where a prime denotes derivative with respect to $R$, $\mbf{D} = \text{diag}(1,~ 1, ~1, ~\mcal{D}_1)$, $\mbf{y}' = [{r^{(0)}}'~ {r^{(0)}}''~ {r^{(0)}}'''~ {r^{(0)}}^{\text{iv}}]^{T} =  [y_1'~ y_2'~ y_3'~ y_4']^{T}$,  $\mbf{q} = [y_2~ y_3~ y_4~ \mcal{D}_2]^{T}$, and
 $\mcal{D}_1$ and $\mcal{D}_2$ are given as
\begin{align*}
\mcal{D}_1  =~ & 2 \rho_1^3 \rho_2^2 \rho^6 \bar{h}^2 \lm^6 \bigg[ 4 \rho^2 \lm^6 + \rho_1^2 \rho_2^4 \bigg],\\
\mcal{D}_2  = ~& 80\,\bar{h}^{2}{\lambda}^{12}{\rho}^{8}{\rho_{1}}^{3}\rho_{2}\,\rho_{3}\, \rho_{4}-72\,\bar{h}^{2}{\lambda}^{12}{\rho}^{8}{\rho_{1}}^{3}{\rho_{3}}^{3}+24\, \bar{h}^{2}{\lambda}^{12}{\rho}^{8}{\rho_{1}}^{2}{\rho_{2}}^{3}\rho_{4}+64\, \bar{h}^{2}{\lambda}^{12}{\rho}^{8}{\rho_{1}}^{2}{\rho_{2}}^{2}{\rho_{3}}^{2} \\
& +82\,\bar{h}^{2}{\lambda}^{12}{\rho}^{8}\rho_{1}\,{\rho_{2}}^{4}\rho_{3}-6\,\bar{h}^{2}{\lambda}^{12}{\rho}^{8}{\rho_{2}}^{6}-32\,\bar{h}^{2}{\lambda}^{12}{\rho}^{7}{\rho_{1}}^{3}{\rho_{2}}^{2}\rho_{4}-24\,\bar{h}^{2}{\lambda}^{12}{\rho}^{7}{\rho_{1}}^{3}\rho_{2}\,{\rho_{3}}^{2} \\
& -200\,\bar{h}^{2}{\lambda}^{12}{\rho}^{7}{\rho_{1}}^{2}{\rho_{2}}^{3}\rho_{3} - 82\,\bar{h}^{2}{\lambda}^{12}{\rho}^{7}\rho_{1}\,{\rho_{2}}^{5}+144\,\bar{h}^{2}{\lambda}^{12}{\rho}^{6}{\rho_{1}}^{3}{\rho_{2}}^{2}\rho_{3}+136\,\bar{h}^{2}{\lambda}^{12}{\rho}^{6}{\rho_{1}}^{2}{\rho_{2}}^{4}\\
& -24\,\bar{h}^{2}{\lambda}^{6}{\rho}^{6}{\rho_{1}}^{5}{\rho_{2}}^{5}\rho_{3}\,\rho_{4}+ 80 \, \bar{h}^{2}{\lambda}^{6}{\rho}^{6}{\rho_{1}}^{5}{\rho_{2}}^{4}{\rho_{3}}^{3}-8\,\bar{h}^{2}{\lambda}^{6}{\rho}^{6}{\rho_{1}}^{4}{\rho_{2}}^{7} \rho_{4}+86\,\bar{h}^{2}{\lambda}^{6}{\rho}^{6}{\rho_{1}}^{4}{\rho_{2}}^{6}{\rho_{3}}^{2}\\
& + 57\,\bar{h}^{2}{\lambda}^{6}{\rho}^{6}{\rho_{1}}^{3}{\rho_{2}}^{8}\rho_{3} + 20\,\bar{h}^{2}{\lambda}^{6}{\rho}^{6}{\rho_{1}}^{2}{\rho_{2}}^{10}+9\,{\lambda}^{8}{\rho}^{6}{\rho_{1}}^{5}{\rho_{2}}^{6}\rho_{3}+9\,{\lambda}^{8}{\rho}^{6}{\rho_{1}}^{4}{\rho_{2}}^{8}-48\,\bar{h}^{2}{\lambda}^{12}{\rho}^{5}{\rho_{1}}^{3}{\rho_{2}}^{3} \\
&-4\,\bar{h}^{2}{\lambda}^{6}{\rho}^{5}{\rho_{1}}^{5}{\rho_{2}}^{6}\rho_{4}-56\,\bar{h}^{2}{\lambda}^{6}{\rho}^{5}{\rho_{1}}^{5}{\rho_{2}}^{5}{\rho_{3}}^{2}-36\,\bar{h}^{2}{\lambda}^{6}{\rho}^{5}{\rho_{1}}^{4}{\rho_{2}}^{7}\rho_{3} +9\, \bar{h}^{2}{\lambda}^{6}{\rho}^{5}{\rho_{1}}^{3}{\rho_{2}}^{9}-9\,{\lambda}^{8}{\rho}^{5}{\rho_{1}}^{5}{\rho_{2}}^{7} \\
& +12\,\bar{h}^{2}{\lambda}^{6}{\rho}^{4}{\rho_{1}}^{6}{\rho_{2}}^{5}\rho_{4}-60\,\bar{h}^{2}{\lambda}^{6}{\rho}^{4}{\rho_{1}}^{6}{\rho_{2}}^{4}{\rho_{3}}^{2}-104\,\bar{h}^{2}{
\lambda}^{6}{\rho}^{4}{\rho_{1}}^{5}{\rho_{2}}^{6}\rho_{3} - 93\,\bar{h}^{2}{\lambda}^{6}{\rho}^{4}{\rho_{1}}^{4}{\rho_{2}}^{8} \\
&+3\,{\lambda}^{2}{\rho}^{4}{\rho_{1}}^{7}{\rho_{2}}^{10}\rho_{3}+92\,\bar{h}^{2}{\lambda}^{6}{\rho}^{3}{\rho_{1}}^{6}{\rho_{2}}^{5}\rho_{3}+64\,\bar{h}^{2}{\lambda}^{6}{\rho}^{3}{\rho_{1}}^{5}{\rho_{2}}^{7}+3\,{\lambda}^{2}{\rho}^{3}{
\rho_{1}}^{7}{\rho_{2}}^{11}+\bar{h}^{2}{\rho}^{2}{\rho_{1}}^{7}{\rho_{2}}^{10}\rho_{3} \\
&-3\,{\lambda}^{2}{\rho}^{2}{\rho_{1}}^{8}{\rho_{2}}^{10}+\bar{h}^{2}\rho\,{\rho_{1}}^{7}{\rho_{2}}^{11}-{\rho_{1}}^{8}{\rho_{2}}^{10}\bar{h}^{2}.
 \end{align*}

 While deriving the above, we have used the dimensionless quantities
 \begin{align}
 \rho = \frac{R}{B} , \quad \rho_1 = \frac{r^{(0)}}{B}, \quad \rho_2 = {r^{(0)}}', \quad \rho_3 = {r^{(0)}}'' B, \quad \rho_4 = {r^{(0)}}''' B^2, \quad \text{and} \quad \bar{h} = \frac{h}{B}. \label{non_dim_base_sol}
 \end{align}
The principal solution for $\lm > 1$ allows the contraction of inner radius and expansion of outer radius of the plate which are investigated for two different boundary conditions. In the first boundary condition, we consider the inner boundary 
of the plate to be unconstrained or free to contract due to growth  ($\lambda>1$) and the outer boundary is constrained (IFOC\footnote{IFOC -- Inner boundary of the plate is unconstrained (free) and outer boundary of the plate is constrained.}).
This condition is inspired from the behaviour of soft biological tissue such as skin where the wounded skin grows to close a wound
\citep{swain2015interfacial, bowden2016morphoelastic}.
The second boundary condition models a constrained inner edge and unconstrained outer edge (ICOF\footnote{ICOF -- Inner boundary of the plate is constrained and the outer boundary of the plate is free.}) of a plate where only the outer edge is allowed to deform  during growth process. 
This condition is akin to the deformation of plants and soft polymeric material such as swollen gels \citep{mora2006buckling, liu2013pattern}.
\subsubsection{Case 1: Constrained outer boundary and unconstrained inner boundary} 
If the inner edge (at $\rho = A/B = A^*$) of the plate is free to contract or expand then the radial stress at inner edge on bottom and top surface is $\left. P_{Rr} \right|_{Z=0} = \left. P_{R r} \right|_{Z = 2h} = 0$ (which corresponds to  $P^{(0)}_{Rr}$ = $P^{(2)}_{Rr} = 0$ using \eqref{decom_plate_eq_1}) and can further be rewritten in terms of dimensionless variables as
\begin{subequations} \label{inner_free_bound_cond}
\begin{align}
& \lambda \bigg[\frac{2 \rho_1}{\lambda} - \frac{2 \lambda^5 {A^*}^2} {\rho_1^2 \rho_2^3} \bigg] = 0,  \\
& 8 \rho_2^2 {A^*}^5 \lm^6 \rho_3 + 4 \rho_2 \rho_1 \rho_4 {A^*}^5 \lm^6 - 4 \rho_3^2 \rho_1 {A^*}^5 \lm^6 - 8 \rho_2 ^3 {A^*}^4 \lm^6- 4 \rho_2 \rho_1 \rho_3 {A^*}^4 \lm^6 + 2 \rho_2^8 \rho_1 {A^*}^3  \nonumber \\
& \quad + 7 \rho_2^6 \rho_1^2 {A^*}^3 \rho_3  + \rho_2^5 \rho_1^3 \rho_4 {A^*}^3 + 8 \rho_1^4 \rho_1^3 \rho_3^2 {A^*}^3 + 8 \rho_2^2 \rho_1 {A^*}^3 \lm^6 + 3 \rho_2^7 \rho_1^2 {A^*}^2 - \rho_2^5 \rho_1^3 \rho_3 {A^*}^2 - 15 \rho_2^6 \rho_1^3 {A^*}  \nonumber\\
& \quad - 6 \rho_2^4 \rho_1^4 \rho_3 {A^*} + 10 \rho_2^5 \rho_1^4  = 0. 
\end{align}
\end{subequations}
If the outer edge of the plate is constrained then the displacement of bottom and top surface at $\rho = 1$ is 0 i.e., $r^{(0)}(B) = B$, and $r^{(0)} + (2h) r^{(1)} + 2h^2 r^{(2)} = B$ 
and is given by
\begin{subequations} \label{outer_const_bound_cond}
\begin{align}
\rho_1(1) &= 1, \\
{r^{(2)}}(1) & = -2 \rho_2^2 \lm^6 - 2 \rho_1 \rho_3 \lm^6 + 2 \rho_2 \rho_1 \lm^6 - \rho_2^4 \rho_1^3 \rho_3 - \rho_2^5 \rho_1^3 + \rho_2^4 \rho_1^4 = 0. 
\end{align}
\end{subequations}

\subsubsection{Case 2: Constrained inner boundary and unconstrained outer boundary}
If the inner edge of the plate is constrained then the displacement of bottom and top surface at the inner edge $ \big(\left. \rho \right|_{A^*} = 0 \big)$ is 
\begin{subequations} \label{eq:inner_const_sec_bound}
\begin{align}
r^{(0)}(A^*) & = A^* \to \rho_1(A^*) = A^{*},\\
{r^{(2)}}(A^*) & = -2 \rho_2^2 {A^{*}}^4 \lm^6 - 2 \rho_1 \rho_3 {A^{*}}^4  \lm^6 + 2 \rho_2 \rho_1  {A^{*}}^3 \lm^6 - \rho_2^4 \rho_1^3 \rho_3 {A^{*}}^2 - \rho_2^5 \rho_1^3 {A^{*}} + \rho_2^4 \rho_1^4 = 0.
\end{align}
\end{subequations}
If the outer edge is unconstrained, the radial stress $P^{(0)}_{Rr}$ = $P^{(2)}_{Rr} = 0$ at $\rho  = 1$ is expressed as
\begin{subequations} \label{eq:outer_free_inner_const}
\begin{align}
& \lambda \bigg[\frac{2 \rho_1}{\lambda} - \frac{2 \lambda^5} {\rho_1^2 \rho_2^3} \bigg] = 0, \\
& 8 \rho_2^2 \lm^6 \rho_3 + 4 \rho_2 \rho_1 \rho_4  \lm^6 - 4 \rho_3^2 \rho_1  \lm^6 - 8 \rho_2 ^3  \lm^6- 4 \rho_2 \rho_1 \rho_3  \lm^6 + 2 \rho_2^8 \rho_1 + 7 \rho_2^6 \rho_1^2  \rho_3 + \rho_2^5 \rho_1^3 \rho_4  \nonumber \\
& \qquad  + 8 \rho_1^4 \rho_1^3 \rho_3^2  + 8 \rho_2^2 \rho_1 \lm^6 + 3 \rho_2^7 \rho_1^2  - \rho_2^5 \rho_1^3 \rho_3  - 15 \rho_2^6 \rho_1^3 - 6 \rho_2^4 \rho_1^4 \rho_3 + 10 \rho_2^5 \rho_1^4 = 0.
\end{align}
\end{subequations}
\subsubsection{Numerical pre-buckling solution }
\begin{figure}
\centering
\includegraphics[width=0.9\linewidth]{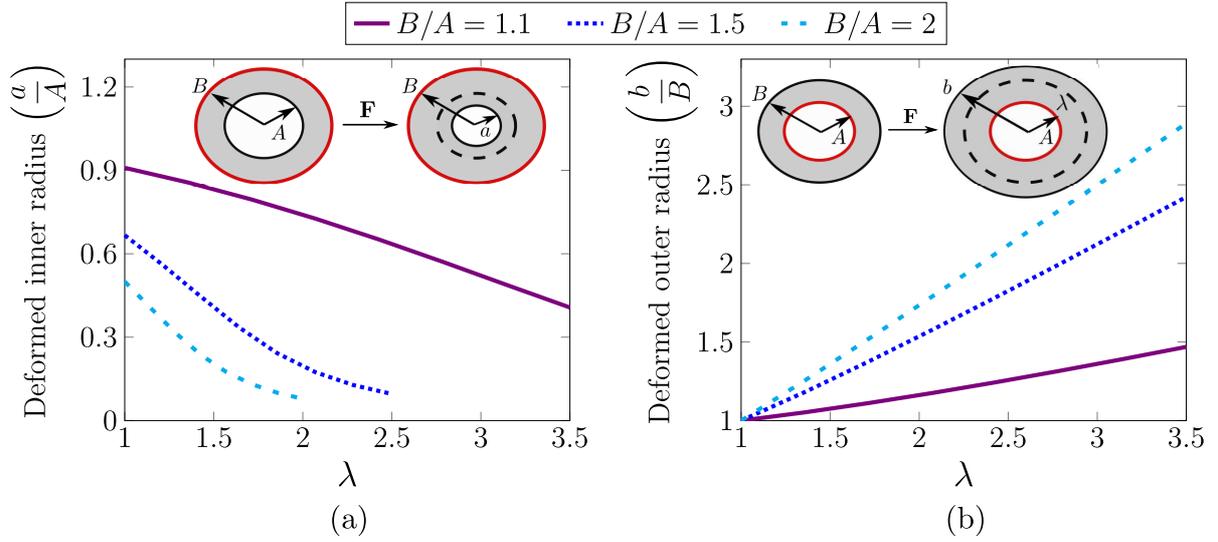}
\caption{a) Dependence of the deformed inner radius ($a/A$) on growth factor $\lambda$ at $\bar{h} = 0.03$ when outer edge of the plate is constrained and inner edge is unconstrained (IFOC), and b) Dependence of deformed outer radius ($b/B$) on growth factor $\lambda$ at $\bar{h} = 0.03$ when the inner edge of the plate is constrained and the outer edge is unconstrained (ICOF). In the first IFOC case, the inner boundary of the plate contracts and in the second ICOF case the outer unconstrained boundary expands on the application of $\lambda$.  The deformed radius for both the cases is plotted for three different values of radius ratio that are $B/A = 1.1, ~ 1.5,~ \text{and}~ 2$.}  \label{fig:radius_to_lam_base_state_sol}
\end{figure}

In this section, we discuss the deformation associated with the principal solution \eqref{principal_sol} for both the boundary conditions. In the first case (IFOC), 
we numerically solve the system of ODEs \eqref{eq:sys_ODES_base} subjected to the boundary conditions \eqref{inner_free_bound_cond} and \eqref{outer_const_bound_cond}. The numerical solutions for primary in-plane deformation of an annulus plate for different radius ratios $B/A = 1.1,~ 1.5,~ 2$ (see Figure \ref{fig:radius_to_lam_base_state_sol}) are obtained using the \texttt{bvp4c} solver available in Matlab.
The dependence of deformed inner radius ($a/A$) on the growth parameter ($\lambda>1$) for a plate of thickness $\bar{h} = 0.03$ is presented in Figure \ref{fig:radius_to_lam_base_state_sol}a. 
During growth, the inner boundary of the annular plate contracts resulting in a decrease of the inner radius ($a < A$). The plate with a high
radius ratio $B/A = 2$ shows more contraction even for a smaller value of $\lambda$ compared to the other ratios.
The numerical pre-buckling solution for the second case (ICOF)
is obtained by solving the system of ODEs \eqref{eq:sys_ODES_base} subjected to the boundary conditions \eqref{eq:inner_const_sec_bound} and \eqref{eq:outer_free_inner_const}.
In this case, we have shown the variation of the deformed outer radius ($b/B$) with growth factor $\lambda$ in Figure \ref{fig:radius_to_lam_base_state_sol}b. The outer edge of the plate expands as $\lambda$ increases and  plates with higher radius ratio show more expansion in comparison to plates with smaller radius ratio.
\subsection{Constrained growth of a circular ring} \label{sec:comp_circular_ring}
\begin{figure}
    \centering
    \includegraphics[width = 0.9\linewidth]{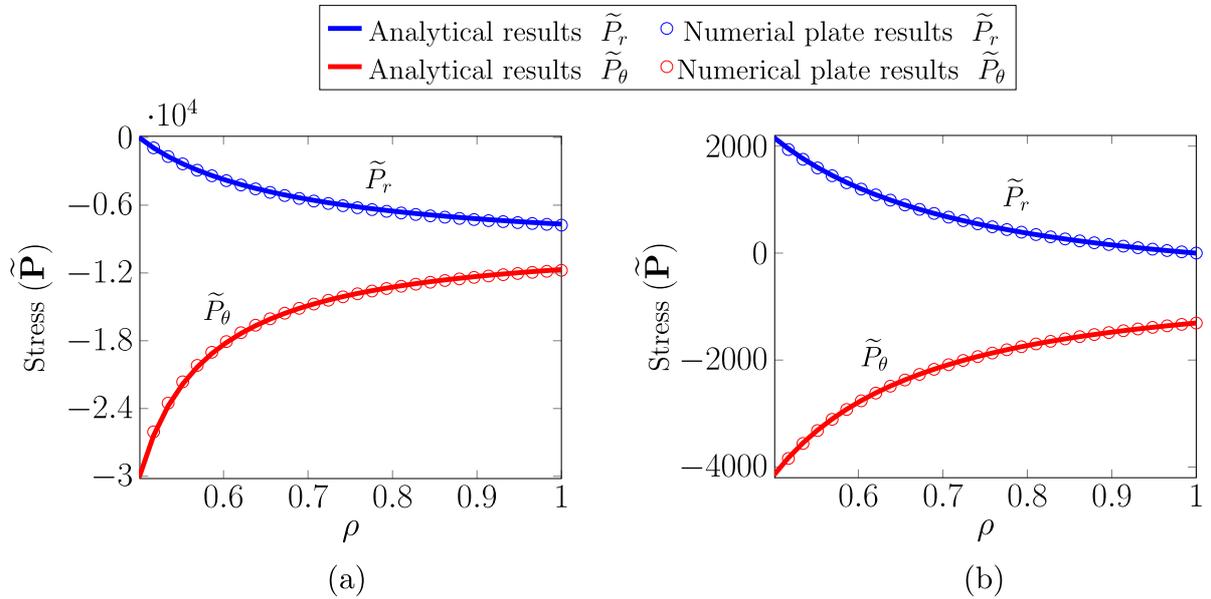}
    \caption{ Comparison of stress distribution obtained using current plate theory with the analytical results for a circular ring \citep{liu2014nonlinear} for the both a) IFOC, and b) ICOF boundary conditions. The Piola Kirchhoff stress (both radial $\widetilde{P}_r$ and circumferential $\widetilde{P}_\theta$ stress) are plotted against the normalised radius $\rho$. 
    The radial and circumferential stress distribution curves are shown by blue and red colour lines, respectively. Numerical results of the current model are in perfect agreement with the analytical results.}
    \label{fig:annulus_base_state_valid}
\end{figure}

To validate the 2-D plate framework, we compare the pre-buckling solution of an isotropically growing thin annular plate with the analytical solution of an incompressible neo-Hookean circular ring growing with planar constant growth i.e., $\lm_{rr} = \lm_{\theta \theta} = 1.35$ provided by \cite[Figure 9]{liu2014nonlinear}.
The ring is subjected to i) outer constrained boundary (similar to IFOC in this work), and ii) inner constrained boundary (similar to ICOF) boundary conditions. The stress distribution curves for i), and ii) boundary conditions are plotted in Figures \ref{fig:annulus_base_state_valid}a, and \ref{fig:annulus_base_state_valid}b, respectively.
A numerical solution of the incompressible circular ring with same boundary condition using solid-shell based finite element approach subjected to identical isotropic growth function is also given by \cite{zheng2019solid}.
Both the analytical and numerical solution are in good agreement with each other.
To test the accuracy of current plate theory, we numerically solved \eqref{eq:sys_ODES_base} subjected to both IFOC (\eqref{inner_free_bound_cond} - \eqref{outer_const_bound_cond}) and ICOF (\eqref{eq:inner_const_sec_bound} - \eqref{eq:outer_free_inner_const}) boundary conditions using same parameters as plate thickness, $2 h =  0.001 \rightarrow 2 \bar{h} = 0.0005$, growth stretch $\lm = 1.35$ and ground state shear modulus $2C_0 = 4000$ by using the \texttt{bvp4c} solver in Matlab.
The comparison of current numerical results with the existing analytical results are in good agreement.
As shown in Figure \ref{fig:annulus_base_state_valid}, numerical results obtained using the current theory are in agreement with the analytical results of \cite{liu2014nonlinear} and numerical results of \cite{zheng2019solid}.

\section{Linear bifurcation analysis} \label{sec:Bifurcation_analysis}
We derive the PDEs for the onset of buckling in an isotropically growing annular plate.
We seek a bifurcation solution close to the primary solution by using three different types of perturbations: (a) asymmetric perturbation along radial, circumferential and thickness directions (i.e, $r,\theta, z$ coordinates), (b) axisymmetric perturbation along radial and thickness directions (i.e, $r, z$ coordinates), and (c) asymmetric perturbation along circumferential and thickness directions (i.e, $\theta, z$ coordinates). 
\subsection{Perturbation along radial, circumferential and thickness direction} \label{sec:gen_perturb}

Consider the following small asymmetric perturbations to the pre-buckling solution \eqref{principal_sol} scaled by a parameter $0<\epsilon \ll 1$ 
\begin{align} 
{r}^{(0)}({R,\Theta}) &= r(R) + \epsilon \Delta U(R) \cos(m \Theta) ,\nonumber\\
\theta^{(0)}(R,\Theta) &= \Theta + \epsilon \Delta V(R) \sin(m \Theta) ,\label{ansatz4}\\
z^{(0)}(R,\Theta) &= C_z + \epsilon \Delta W(R) \cos(m\Theta), \nonumber
\end{align}
where $m=1,\ 2, \ 3, \ ...$ represents the wave number in the circumferential direction and $C_z$ is a constant that models rigid motion of the plate.
Upon substituting \eqref{ansatz4} in the plate governing equation \eqref{mod_gov_2-D_plate_eq}, we obtain ODEs in terms of dimensionless displacement functions $ U$, $V$ and $W$ as
\begin{subequations} \label{general_perturb_eq}
\begin{align}
&a_1 U'' + a_2 U' + a_3 U + a_4 V' + a_5  V + a_6  W''' + a_7  W'' + a_8  W' + a_9  W = 0, \label{general_perturb_eq1}  \\
&b_1 V'' + b_2  V' + b_3 V + b_4  U' + b_5  U +  b_6  W'' + b_7  W' + b_8  W = 0, \label{general_perturb_eq2}\\
&c_1  W'' + c_2  W' + c_3  W + c_4  U''' + c_5  U'' + c_6 U' + c_7  U + c_8  V'' + c_9  V' + c_{10}  V \nonumber \\
& \qquad \qquad \qquad \qquad \qquad  \qquad + c_{11}  W^{\text{iv}} + c_{12}  W''' + c_{13}  W'' + c_{14}  W' 
+ c_{15}  W = 0, \label{general_perturb_eq3}
\end{align}
\end{subequations}
where  $ U = \dl U/ B, ~ V = \dl V/B,$ and $ W = \dl W/B$. 
Equations \eqref{general_perturb_eq1} -- \eqref{general_perturb_eq3} are rewritten into a system of  first order differential equations by substituting 
$[y_1,\ y_2,\  ..., \ y_8] = [ U,\   U',\   V,\  V',\   W,\  W',\  W'',\  W''' ]$
resulting in
\begin{align}
\mbf{HY}' = \mbf{g} . \label{gen_sys_of_eqs}
\end{align} 
Here, $\mbf{H} = \text{diag} (1,~a_1,~ 1,~b_1,~1,~1,~1,~c_{11})$ is an $8 \times 8$ matrix, and $\mbf{Y}' = [y_1',~ y_2',~ y_3',~ y_4',~ y_5',~ y_6',~ y_7',~ y_8']^T$, and $\mbf{g} = [y_1,~ y_2,~ y_3,~ y_4,~ y_5,~ y_6,~ y_7,~ y_8]^T$ are $8 \times 1$ column vectors.
The coefficients in \eqref{general_perturb_eq1} -- \eqref{general_perturb_eq3} are detailed in the supplementary document.
The displacement boundary conditions for the IFOC case are
\begin{subequations} \label{bound_cond_gen_perturb}
\begin{align}
&U'(A^*) = V(A^*) =  W''(A^*) =   W'''(A^*) = 0, \label{eq:IFOC_a}\\
& U(1) =  V(1) = W(1) =  W'(1) = 0, \label{eq:IFOC_b}
\end{align}
\end{subequations}
and for the ICOF case are
\begin{subequations} \label{eq:bc_gen_perturb_ICOF}
\begin{align}
&U(A^*) = V(A^*) =  W(A^*) =   W'(A^*) = 0, \label{eq:ICOF_a}\\
& U'(1) =  V(1) = W''(1) =  W'''(1) = 0. \label{eq:ICOF_b}
\end{align}
\end{subequations}

Here, equations \eqref{eq:IFOC_a} (respectively, \eqref{eq:ICOF_b}) correspond to unconstrained inner edge (respectively, outer edge) of annular plate which allows contraction (respectively, expansion) and no restriction of bending moment ($ W'' = 0$) and transverse shear ($ W''' = 0$). 
Equations \eqref{eq:IFOC_b} and \eqref{eq:ICOF_a} correspond to clamped (constrained) outer and inner edges of the plate, respectively.

\subsubsection{Perturbation along radial and thickness direction} \label{sec:R-Z_perturbation}
In this case, the linear bifurcation analysis is performed by perturbing the principal solution \eqref{principal_sol} axisymmetrically with a small parameter $0<\epsilon \ll 1$ by using the following ansatz
\begin{align}
{r}^{(0)}({R},\Theta) = r(R) + \epsilon \Delta U(R) \cos(m \Theta), \qquad
\theta^{(0)} = \Theta, \qquad
z^{(0)}(R,\Theta) = C_z + \epsilon \Delta W(R) \cos(m\Theta), \label{R-Z-pertrb_ansatz}
\end{align}
where `$m$' is again an integer  representing the wave number in the circumferential direction. 
On substituting \eqref{R-Z-pertrb_ansatz} in the plate governing equation \eqref{mod_gov_2-D_plate_eq}, we obtain ODEs in terms of the dimensionless displacement functions $U$ and $W$ which one can simply obtain by setting the coefficients $a_4 = a_5 =0$ in \eqref{general_perturb_eq1} and $c_8 = c_9 = c_{10} = 0$ in \eqref{general_perturb_eq3}. 
The system \eqref{general_perturb_eq} is now reduced to two ODEs which is further simplified into six first-order linear differential equations of the form
\begin{align}
\mbf{Bm}' = \mbf{m}, \label{eq:R_Z_pertrb_sys}
\end{align}
where $\mbf{m} = [U, ~ U', ~W, ~  W', ~  W'', ~  W''']$ and $\mbf{B} = \text{diag}(1,~ a_1,~ 1,~ 1,~ 1, ~ c_{11})$. 
The IFOC boundary condition  is given as
\begin{align}
 U'(A^*) =  W''(A^*) =  W'''(A^*) = 0, \quad \text{and} \quad 
 U(1) =  W(1) =  W'(1) = 0, \label{eq:R_Z_IFOC}
\end{align}
and the ICOF condition  associated with this type of perturbation yields
\begin{align}
 U(A^*) =  W(A^*) =  W'(A^*) = 0, \quad \text{and} \quad 
 U'(1) =  W''(1) =  W'''(1) = 0. \label{eq:R_Z_ICOF}
\end{align}

\subsubsection{Perturbation along circumferential and thickness direction} \label{sec:theta-Z perturb}
In this case, the bifurcation solution is obtained by applying an asymmetric perturbation to the principal solution  along circumferential and thickness directions using the following ansatz
\begin{align}
{r}^{(0)}({R}) = r(R) , \quad
\theta^{(0)}(R,\Theta) = \Theta + \epsilon \Delta V(R) \sin(m \Theta),~ \text{and} ~
z^{(0)}(R,\Theta) = C_z + \epsilon \Delta W(R) \cos(m\Theta), \label{Theta-Z_ansatz}
\end{align}
where $m=1,2,3 \ ...$ represents the circumferential wave number. 
In this type of perturbation, the incremental elastic strain is dependent on the $\Theta \text{-}Z$ coordinates. Thus,
the incremental ODEs are obtained in terms of dimensionless displacement functions $V$ and $W$ by substituting \eqref{Theta-Z_ansatz} in \eqref{mod_gov_2-D_plate_eq} or we can obtain these ODEs by setting the coefficients $b_4 = b_5 = 0$ in \eqref{general_perturb_eq2} and $c_4 = c_5 = c _6 = c_{7} = 0$ in \eqref{general_perturb_eq3}.
The resulting three equations in the system \eqref{gen_sys_of_eqs} is again reduced to two ODEs which further can be rewritten in the form of 
\begin{align}
\mbf{K t}' = \mbf{t} \label{eq: theta_Z_system}
\end{align} 
where $\mbf{t} = [V, ~ V', ~W, ~  W', ~  W'', ~  W''']$ and $\mbf{K} = \text{diag}(1,~ b_1,~ 1,~ 1,~ 1, ~ c_{11})$.
The IFOC boundary conditions for this case are
\begin{align}
 & V (A^*) = W'' (A^*) = W'''(A^*) = 0,  \quad \text{and} \quad
V (1) = W(1) = W'(1) = 0.  \label{eq:theta_Z_IFOC}
\end{align}
The ICOF boundary conditions are given by
\begin{align}
& V (A^*) = W(A^*) = W'(A^*) = 0,  \quad \text{and} \quad
V(1) = W''(1) = W'''(1) = 0.  \label{eq:theta_Z_ICOF}
\end{align}
\section{Results} \label{sec:Result_sec}
Before presenting the bifurcation solutions for hyperelastic annular plates, we compare the bifurcation results obtained by the current  theory with the FvK plate theory in the small thickness regime for validation.
\subsection{Growth-induced instability in a rectangular plate} \label{sec:FvK_shell_comp}
\begin{figure}
    \centering
    \includegraphics[width = 0.9\linewidth]{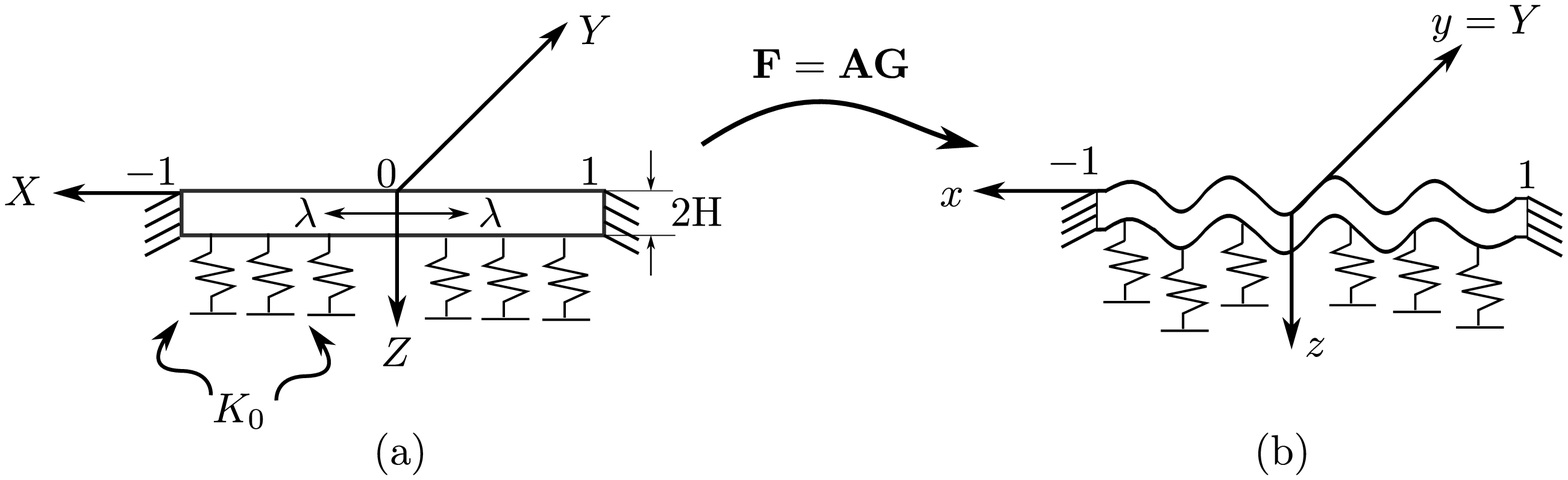}
    \caption{Possible mode of buckling of a rectangular plate under growth-induced deformation. The plate is in a plane strain condition in the $Y$-direction and subjected to Winkler support with an effective stiffness, $K_0$. 
    The reference configuration (a) transforms to the deformed configuration (b) at the critical value of growth factor i.e., $\lambda = \lambda_{cr}$. 
    Compression due to a constrained boundary leads to buckling of the plate.}
    \label{fig:rec_plate_valid}
\end{figure}

Consider a rectangular plate of thickness $2H$ clamped at the ends $X = \pm 1$ (as shown in Figure \ref{fig:rec_plate_valid}) subjected to a plane strain in the $Y$-direction and a Winkler foundation with effective stiffness $K_0$ at $Z=-2H$.
Analytical results for this problem are derived by \cite{wang2018consistent}.
The Winkler foundation allows for analytical solution \citep{dervaux2010localized}.
The plate is made up of an incompressible neo-Hookean material growing under uni-axial constant growth field for which the growth tensor takes the form $\mbf{G} = \text{diag}(\lambda,~ 1,~ 1)$. 
The Winkler support provides traction only in the transverse direction at the bottom surface of the plate given as $t_3 = -K_0 \lm W_0 = -K_0 \lm \big[ z^{(0)}/[2\bar{H}] + z^{(1)} +  \bar{H} z^{(2)} + 2/3 [\bar{H}]^2 z^{(3)} - 1 \big] $ where $\bar{H}$ is the ratio of plate thickness to length (in the X direction) i.e., $\bar{H} = H/L$, $K_0$ is the elastic constant of the foundation and $W_0$ is the transverse component of the displacement.
We perturb the principal solution scaled by a small parameter  $\epsilon$; $x^{(0)}(X) = X_0 + \epsilon  U(X)$,~ $z^{(0)}(X) = -2 \bar{H} (\lm - 1) + \epsilon W(X)$ where $x^{(0)}$, $z^{(0)}$ are the deformed coordinates, and $U$ and $W$ are the displacements in axial and transverse direction, respectively.
Upon substituting this ansatz in \eqref{mod_gov_2-D_plate_eq}, equating \eqref{mod_gov_transverse_plate_eq} to $t_3$, and collecting only $O(\epsilon)$ terms,
the governing plate differential equation is given by
\begin{align}
    \eta_1 W' + \eta_2 W'' + \eta_3  W^{\text{iv}} = 0, \label{eq:rec_plate}
\end{align}
subjected to clamped boundary conditions, $W'(\pm 1) = W'''(\pm 1) =0$.
Here
\begin{align*}
    \eta_1  & = -\frac{\alpha \lm}{2 \bar{H}},\\
    \eta_2 &  = \frac{2}{\lm} [1-\lm^4] - \bar{H} \alpha  \lm^3 + \frac{2 \bar{H} \alpha \lm^3 }{1 + 3\lm^4} \big  [1 + \lm^4 \big],\\
    \eta_3 & = -\frac{4 \bar{H}^2 (1 + \lm^4)}{ (1 + 3 \lm^4)} \bigg[ 2 \lm^5 - 3 \lm^4 + 2 \lm -1 \bigg] - \frac{4}{3} \bar{H}^2 [1 + \lm^4] - \frac{4 \bar{H}^3 \alpha \lm}{3 [1 + 3 \lm^4]} [1 + 2 \lm^4 ] [1+ \lm^4], 
\end{align*}
and $\alpha = K_0/C_0$ where $2 C_0 = \mu$ is the ground state shear modulus of the hyperelastic material.
We follow the same numerical scheme as described in Section \ref{sec:Numerical_scheme} and obtained the bifurcation solution by numerically solving the GDE \eqref{eq:rec_plate} for critical value of growth ($\lm_{cr})$ responsible for the onset of instability. 
We compare our numerical results with the existing analytical results using current theory and numerical result using FvK plate theory provided by \cite{wang2018consistent} in Table \ref{tab:comp_results_FvK}. The current numerical results are an almost exact match with the analytical solution and very close to the results obtained by the FvK theory.
\begin{table}
\caption{Comparison of bifurcation solution using finite strain plate theory with analytical and numerical FvK plate results.
 The value of critical growth parameter $\lm_{cr}$ is compared for the plate thickness $\bar{H} = 0.015$ and associated mode numbers $(n)$. The results are in good agreement with analytical results and close to the FvK results.}
 \vspace{0.1in}
    \centering
    \begin{tabular}{c|c|c|c|c}
    \hline
    \hline
      Mode number   &  $n=7$ & $n=8$ & $n=5$ & $n = 9$ \\
      \hline
       FvK theory   & 1.01596 & 1.01712 & 1.01814 & 1.01916 \\
       Analytical results \citep{wang2018consistent} & 1.01626 & 1.01751 & 1.01839 & 1.01969 \\
    Current results & 1.01626 & 1.01751 & 1.01839 & 1.01969\\
       \hline
    \end{tabular}
    \label{tab:comp_results_FvK}
\end{table}

\subsection{Results for annular plate }
The previous Section \ref{sec:Bifurcation_analysis} discussed the possible type of perturbations (axisymmetry and asymmetric), corresponding ODEs and boundary conditions for the stability analysis. 
In this section, we numerically solve the derived ODEs to determine the critical growth factor at the onset of bifurcation. The numerical strategy used to derive the bifurcation solution is detailed in Section \ref{sec:Numerical_scheme}. We have shown only one set of bifurcation solutions in Subsection \ref{sec:Bifurcation_sol_gen_perturb} when a plate is perturbed in $R \text{-} \Theta \text{-} Z$ direction and subjected to IFOC boundary condition. This is to illustrate the stability of buckling solution of the growing plate when perturbed. Then in Subsection \ref{sec:Comp_of_bif_sol}, we compare the bifurcation solution of each type of perturbation considered in Section \ref{sec:Bifurcation_analysis} and investigate the type of perturbation that results in the energetically preferred bifurcation solution. In the next Section \ref{sec:buckling_sol_diff_bound}, we have shown the comparison between the preferred bifurcation solutions for both boundary conditions.
\subsection{Numerical bifurcation analysis of growing annular plate}  \label{sec:Numerical_scheme}
In this section, we numerically solve the ODEs obtained in Section \ref{sec:gen_perturb} for the 
critical growth factor ($\lm_{cr}$)  responsible for the onset of instability. 
The numerical solutions of the resulting boundary value problems (BVPs) are computed using the compound matrix method \citep{haughton1997eversion, mehta2021growth} as well as the standard shooting method or determinant method \citep{haughton1979bifurcation,saxena2018finite}.
Both the numerical methods are implemented in Matlab 2018a and the computed results are witin a very small norm. 
However, the compound matrix method is much faster than the shooting method \citep{mehta2021instabilities}. 
Equations are integrated using the $\texttt{ode45}$  ODE solver that implements an explicit Runga-Kutta method and then the $\texttt{fminsearchbnd}$ optimisation subroutine \citep{JohnDErrico2021} based on a Nelder-Mead simplex algorithm is used to minimise the errors.

\subsubsection{Bifurcation solution for a plate with clamped outer edge condition} \label{sec:Bifurcation_sol_gen_perturb}
\begin{figure}
\centering
\includegraphics[width=0.95\linewidth]{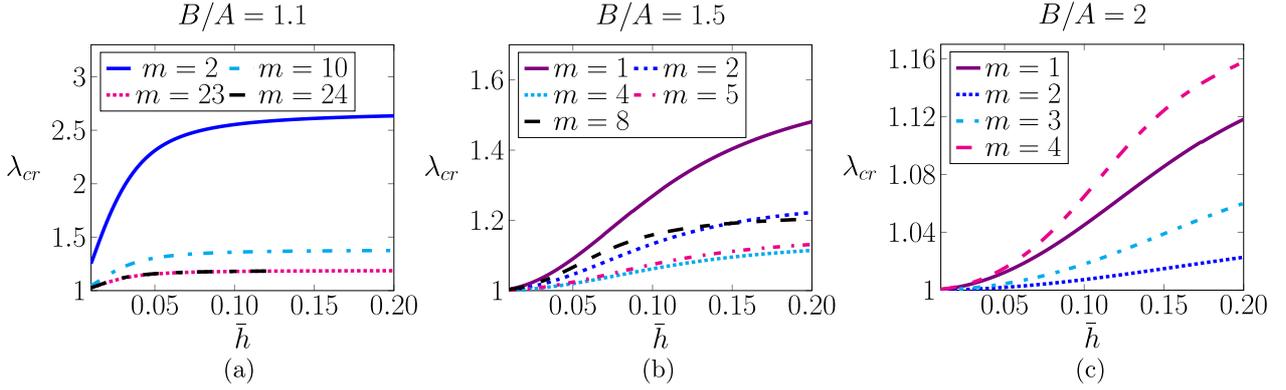}
\caption{Dependence of the critical growth factor on the thickness of the plate ($\bar{h}$) for various aspect ratios a) $B/A = 1.1$, b) $B/A = 1.5$, and c) $B/A = 2$ that results in a perturbation in the radial-hoop-thickness direction. The bifurcation solution in each case is obtained for the IFOC boundary condition.  
Lowest value of $\lambda_{cr}$ in each case corresponds to the minimum energy and therefore the preferable solution and determines the critical wave number $m$ that directly relates to the number of wrinkles in the deformed configuration.}
\label{fig:L_vs_h_gen_perturb}
\end{figure}

In this section, we evaluate the value of $\lm_{cr}$ for the first case as discussed in \ref{sec:gen_perturb} i.e., perturbation along $R \text{-} \Theta \text{-} Z$ direction to demonstrate the behaviour of the bifurcation solution for an annular plate. For this, we numerically solve the system \eqref{gen_sys_of_eqs} subjected to IFOC boundary condition \eqref{bound_cond_gen_perturb}. 
The dependence of $\lm_{cr}$ on the plate thickness ($\bar{h}$) with different radius ratios $B/A = 1.1,~ 1.5,~ 2$ at various circumferential wavenumbers ($m$) is shown in Figure \ref{fig:L_vs_h_gen_perturb}. 
The $\lm_{cr}$ monotonically increases with $\bar{h}$ suggesting that thicker plates with high bending stiffness require more growth to cause wrinkling instability.
Furthermore, the magnitude of $\lm_{cr}$ decreases with the increase of $B/A$ due to an apparent decrease in the boundary layer effects of the annular plate. 
We also observe that the critical wavenumber ($m_{cr}$) depends on the thickness, radius ratio, and boundary layer effects of the plate. Higher modes are energetically preferred 
for the low $B/A$ values
whereas lower modes are stable with an increasing value of $B/A$. 
For a plate with smaller radius ratio, $B/A = 1.1$, the obtained critical wavenumber is $m_{cr} = 22$ in the thin regime ($0.05 < \bar{h} <  0.1$), and $m_{cr} = 23$ in the thick regime ($\bar{h}> 0.1$).
However, for $B/A = 1.5$ and $2$, the critical wavenumbers are $m_{cr} = 4$ and $m_{cr} = 2$, respectively.

\subsection{{Comparison between axisymmetric and asymmetric bifurcation}} \label{sec:Comp_of_bif_sol}
Here, we compare the bifurcation solutions for axisymmetric and asymmetric perturbations as discussed in Section \ref{sec:R-Z_perturbation} and \ref{sec:theta-Z perturb}, respectively. 
The $\lm_{cr}$ is calculated by numerically solving the systems \eqref{eq:R_Z_pertrb_sys} and \eqref{eq: theta_Z_system} subjected to the IFOC boundary conditions \eqref{eq:R_Z_IFOC} and \eqref{eq:theta_Z_IFOC}, respectively using the compound matrix method.  
Table \ref{tab:comp_diff_pertrb} shows the dependence of $\lm_{cr}$ on plate thickness ($\bar{h}$) and plate radius ratio ($B/A$) at the corresponding critical mode number ($m_{cr}$) for each type of the considered perturbation. 
The bifurcation solution corresponding to axisymmetric (i.e., along $R \text{-} Z$) and asymmetric (i.e., along $\Theta \text{-} Z$) perturbations to principal solution \eqref{principal_sol} follow a trend similar to the solution obtained in Section \ref{sec:Bifurcation_sol_gen_perturb}. 
The bifurcation solution corresponding to axisymmetric perturbation results in high $\lm_{cr}$, thus requiring higher growth factor as compared to bifurcation solution associated with asymmetric perturbations.

For  plates with $B/A = 1.5, ~ 2 $, the bifurcation solution obtained for the $\Theta \text{-} Z$ perturbation has the minimum energy (highlighted values in Table \ref{tab:comp_diff_pertrb} correspond to lowest values of $\lm_{cr}$ and therefore the preferred bifurcation solution) suggesting that the asymmetrically perturbed solution is energetically preferred.
On the other hand, for a plate with low radius ratio $B/A = 1.1$, the bifurcation solution
obtained for the $R \text{-} \Theta \text{-} Z$ perturbation is energetically preferred.
However, for small thickness values $\bar{h}<=0.05$, difference in the the critical value $\lm_{cr}$ for each of the perturbation types is very small and it is hard to predict the preferred bifurcation mode.
It is also observed that the critical wave number $m_{cr}$ decreases with an increase in $B/A$ for a fixed value of the plate thickness.
This is expected since thinner rings (low $B/A$) have a larger circumference in relation to their surface area and therefore tend to develop more wrinkles compared to bulkier rings (high $B/A$) and is corroborated by the experimental and analytical findings of \cite{mora2006buckling}  and \cite{liu2013pattern}. 

For $B/A = 1.1$, the critical load $\lambda_{cr}$ and the corresponding $m_{cr}$ increase with the thickness $\bar{h}$ . 
For higher $B/A$ values, $m_{cr}$ remains constant but only $\lm_{cr}$ increases with the thickness $\bar{h}$.
In general higher thickness value $\bar{h}$ results in larger compressive stresses at the bifurcation point due to a larger value of $\lm_{cr}$ which is discussed in the subsequent section. 
As a result, bifurcation occurs with a higher wavenumber $m_{cr}$ for $B/A =1.1$ as $\bar{h}$ increases.
For $B/A =1.5, 2$, the change in $\lm_{cr}$ with the $\bar{h}$ value is not significant and as a result $m_{cr}$ stays constant.
%

We conclude that the $\Theta \text{-} Z $ (asymmetric) type of perturbation is appropriate for the annular plate with moderate and high radius ratio when subjected to IFOC condition as compared to other perturbations. 
In the subsequent section, we analyse the influence of ICOF boundary condition (i.e., inner edge of the plate is clamped) on the bifurcation solutions obtained using all types of considered perturbations and compare these results with the IFOC results.

\begin{sidewaystable}
\centering
\caption{Comparison of dependence of critical growth factor $\lambda_{cr}$ on the dimensionless plate thickness $(\bar{h})$ at the corresponding critical circumferential wavenumber ($m_{cr}$) for plate radius ratio, $B/A = 1.1, ~ 1.5,~ 2$. 
Critical value of the growth factor is obtained for both axisymmetric and asymmetric perturbations to the principal solution subjected to IFOC boundary constraint. 
The highlighted boxed values represent the lowest value of $\lm_{cr}$ and corresponds to the preferred bifurcation solution.} \label{tab:comp_diff_pertrb}
\begin{tabular}{p{1cm}| p{6cm}| p{5.5cm} | p{5.4cm}}
\hline
\hline
\begin{tabular}{c}
$\bar{h}$
\end{tabular}
&
 \begin{tabular}{c}
  $B/A = 1.1$ 
  \\
 \begin{tabular}{p{1.5cm} p{1.3cm} p{1cm}}
 \centering
 \begin{tabular}{c}
 $\lm_{cr}$  \\
 \small{($R \text{-} Z$)}
 \end{tabular}
& 
\begin{tabular}{c}
$\lm_{cr}$\\
(\small{$\Theta \text{-} Z$)}
\end{tabular}
&
\begin{tabular}{c}
$\lm_{cr}$ \\
\small{($R \text{-} \Theta \text{-} Z$)}
\end{tabular}
\end{tabular} 
\end{tabular}
&
 \begin{tabular}{c}
 $B/A = 1.5$ 
 \\
 \begin{tabular}{p{1.2cm} p{1.2cm} p{1cm}}
 \begin{tabular}{c}
$\lm_{cr}$ \\
\small{($R \text{-} Z$)}
\end{tabular}
& 
\begin{tabular}{c}
$\lm_{cr}$ \\
\small{($\Theta \text{-} Z$)}
\end{tabular}
&
\begin{tabular}{c}
$\lm_{cr}$ \\
\small{($R \text{-} \Theta \text{-} Z$)}
\end{tabular}
\end{tabular} 
\end{tabular}
&
 \begin{tabular}{c}
  $B/A = 2$ 
 \\
 \begin{tabular}{p{1.2cm} p{1.2cm} p{1cm}}
 \begin{tabular}{c}
$\lm_{cr}$ \\
\small{($R \text{-} Z$)}
\end{tabular}
& 
\begin{tabular}{c}
$\lm_{cr}$ \\
\small{($ \Theta \text{-} Z$)}
\end{tabular}
&
\begin{tabular}{c}
$\lm_{cr}$ \\
\small{($R \text{-} \Theta \text{-} Z$)}
\end{tabular}
\end{tabular} 
\end{tabular}
\\
\hline
\begin{tabular}{c}
0.03
\end{tabular}
&
\begin{tabular}{p{1.5cm} p{1.5cm} p{1cm}}
\begin{tabular}{p{2cm}}
1.2106\\
(\small{$m_{cr} = 16$})
\end{tabular}
& 
\begin{tabular}{p{2cm}}
1.1154\\
\small{($m_{cr} = 20$)}
\end{tabular}
&
\begin{tabular}{p{2cm}}
\tc{\boxed{{1.1127}}}\\
\small{($m_{cr} = 20$)}
\end{tabular}
\end{tabular}
&
\begin{tabular}{p{1.4cm} p{1.4cm} p{1cm}}
\begin{tabular}{p{2cm}}
1.0077 \\
 \small{($m_{cr} = 4$)}
\end{tabular}
 & 
\begin{tabular}{p{2cm}}
 \tc{\boxed{{1.0074}}}\\
 \small{($m_{cr} = 4$)}
\end{tabular} 
&
\begin{tabular}{p{2cm}}
1.0076\\
 \small{($m_{cr} = 4$)}
\end{tabular}
\end{tabular}
&
\begin{tabular}{p{1.4cm} p{1.4cm} p{1cm}}
\begin{tabular}{p{2cm}}
1.0007\\
 \small{($m_{cr} = 2$)}
\end{tabular}
&
\begin{tabular}{p{2cm}}
{1.0007}\\
 \small{($m_{cr} = 2$)}
\end{tabular}
&
\begin{tabular}{p{2cm}}
1.0007\\
 \small{($m_{cr} = 2$)}
\end{tabular}
\end{tabular}
\\
\hline
%
\begin{tabular}{c}
0.05
\end{tabular}
&
\begin{tabular}{p{1.5cm} p{1.5cm} p{1cm}}
\begin{tabular}{p{2cm}}
1.4604\\
\small{($m_{cr} = 13$)}
\end{tabular}
&
\begin{tabular}{p{2cm}}
1.1619 \\
\small{($m_{cr} = 22$)}
\end{tabular}
&
\begin{tabular}{p{2cm}}
\tc{\boxed{{1.1547}}}\\
\small{($m_{cr}=21$)}
\end{tabular}
\end{tabular}
&
\begin{tabular}{p{1.4cm} p{1.4cm} p{1cm}}
\begin{tabular}{p{2cm}}
1.0214\\
\small{($m_{cr} = 4$)}
\end{tabular}
 & 
\begin{tabular}{p{2cm}}
\tc{\boxed{{1.0191}}}\\
 \small{($m_{cr} = 4$)}
\end{tabular} 
&
\begin{tabular}{p{2cm}}
1.0200\\
\small{($m_{cr} = 4$)}
\end{tabular}
\end{tabular}
&
\begin{tabular}{p{1.4cm} p{1.4cm} p{1cm}}
\begin{tabular}{p{2cm}}
1.0019\\
 \small{($m_{cr} = 2$)}
\end{tabular}
&
\begin{tabular}{p{2cm}}
{1.0019}\\
 \small{($m_{cr} = 2$)}
\end{tabular}
&
\begin{tabular}{p{2cm}}
1.0019\\
 \small{($m_{cr} = 2$)}
\end{tabular}
\end{tabular}
\\
\hline
\begin{tabular}{c}
0.1
\end{tabular}
&
\begin{tabular}{p{1.5cm} p{1.5cm} p{1cm}}
\begin{tabular}{p{2cm}}
1.7893\\
 \small{($m_{cr} = 10$)} 
\end{tabular}
& 
\begin{tabular}{p{2cm}}
1.1901 \\
 \small{($m_{cr} = 23$)}
\end{tabular}
&
\begin{tabular}{p{2cm}}
\tc{\boxed{{1.1793}}}\\
\small{($m_{cr} = 22$)}
\end{tabular}
\end{tabular}
&
\begin{tabular}{p{1.4cm} p{1.4cm} p{1cm}}
\begin{tabular}{p{2cm}}
 1.0810 \\
\small{($m_{cr} = 4$)}
\end{tabular}
 & 
\begin{tabular}{p{2cm}}
\tc{\boxed{{ 1.0551}}}\\
\small{($m_{cr} = 4$)}
\end{tabular} 
&
\begin{tabular}{p{2cm}}
1.0621\\
\small{($m_{cr} = 4$)}
\end{tabular}
\end{tabular}
&
\begin{tabular}{p{1.4cm} p{1.4cm} p{1cm}}
\begin{tabular}{p{2cm}}
1.0075 \\
 \small{($m_{cr} = 2$)}
\end{tabular}
&
\begin{tabular}{p{2cm}}
\tc{\boxed{{1.0063}}}\\
 \small{($m_{cr} = 2$)}
\end{tabular}
&
\begin{tabular}{p{2cm}}
1.0073\\
 \small{($m_{cr} = 2$)}
\end{tabular}
\end{tabular}
\\
\hline
%
\begin{tabular}{c}
0.15
\end{tabular}
&
\begin{tabular}{p{1.5cm} p{1.5cm} p{1cm}}
\begin{tabular}{p{2cm}}
1.8838\\
 \small{($m_{cr} = 9$)}
\end{tabular}
 & 
\begin{tabular}{p{2cm}}
1.1959 \\
 \small{($m_{cr}= 24$)}
\end{tabular} 
&
\begin{tabular}{p{2cm}}
\tc{\boxed{{1.1843}}}\\
 \small{($m_{cr} = 23$)}
\end{tabular}
\end{tabular}
&
\begin{tabular}{p{1.4cm} p{1.4cm} p{1cm}}
\begin{tabular}{p{2cm}}
 1.1477\\
\small{($m_{cr} = 4$)}
\end{tabular}
  & 
  \begin{tabular}{p{2cm}}
\tc{\boxed{{ 1.0814}}} \\
\small{($m_{cr} = 4$)}
  \end{tabular}
  &
  \begin{tabular}{p{2cm}}
  1.0955\\
\small{($m_{cr} = 4$)}
  \end{tabular}
\end{tabular}
&
\begin{tabular}{p{1.4cm} p{1.4cm} p{1cm}}
\begin{tabular}{p{2cm}}
1.0156\\
 \small{($m_{cr} = 2$)}
\end{tabular}
&
\begin{tabular}{p{2cm}}
\tc{\boxed{{1.0113}}}\\
 \small{($m_{cr} = 2$)}
\end{tabular}
&
\begin{tabular}{p{2cm}}
1.0148\\
 \small{($m_{cr} = 2$)}
\end{tabular}
\end{tabular}
\\
\hline
%
\begin{tabular}{c}
0.2
\end{tabular}
&
\begin{tabular}{p{1.5cm} p{1.5cm} p{1cm}}
\begin{tabular}{p{2cm}}
1.9162\\
 \small{($m_{cr} = 9$)}
\end{tabular}
 &   
\begin{tabular}{p{2cm}}
1.1979\\
 \small{($m_{cr} = 24$)}
\end{tabular}  
&
\begin{tabular}{p{2cm}}
\tc{\boxed{{1.1860}}}\\
 \small{($m_{cr} = 23$)}
\end{tabular} 
\end{tabular}
&
\begin{tabular}{p{1.4cm} p{1.4cm} p{1cm}}
\begin{tabular}{p{2cm}}
 1.1849\\
 \small{($m_{cr} = 3$)}
\end{tabular}
  &   
\begin{tabular}{p{2cm}}
\tc{\boxed{{1.0962}}}\\
\small{($m_{cr} = 4$)}
\end{tabular} 
&
\begin{tabular}{p{2cm}}
1.1144\\
\small{($m_{cr} = 4$)}
\end{tabular}
\end{tabular}
&
\begin{tabular}{p{1.4cm} p{1.4cm} p{1cm}}
\begin{tabular}{p{2cm}}
1.0247\\
 \small{($m_{cr} = 2$)}
\end{tabular}
&
\begin{tabular}{p{2cm}}
\tc{\boxed{{1.0155}}}\\
 \small{($m_{cr} = 2$)}
\end{tabular}
&
\begin{tabular}{p{2cm}}
1.0228\\
 \small{($m_{cr} = 2$)}
\end{tabular}
\end{tabular}
\\
\hline
\end{tabular}
\end{sidewaystable}

\subsection{Influence of boundary condition on the buckling solution}
\label{sec:buckling_sol_diff_bound} 
In this section, we evaluate $\lm_{cr}$ for an annular plate considering all type of perturbations by numerically solving \eqref{gen_sys_of_eqs}, \eqref{eq:R_Z_pertrb_sys}, and \eqref{eq: theta_Z_system}
subjected to ICOF conditions \eqref{eq:bc_gen_perturb_ICOF}, \eqref{eq:R_Z_ICOF}, and \eqref{eq:theta_Z_ICOF}, respectively and compare with the solutions obtained in the previous section.
The values of $\lm_{cr}$ are compared for each type of perturbation and then we identify the preferred bifurcation solution by studying the lowest value of $\lm_{cr}$ (at the corresponding $m_{cr}$).
Table \ref{tab: comp_two_bound_at_critical_modes} shows the dependence of preferred bifurcation solution (lowest $\lm_{cr}$) on plate thickness ($\bar{h}$) and radius ratio $B/A$ when the growing plate is subjected to ICOF boundary condition. 
This table also compares the lowest values of $\lm_{cr}$ of the plate subjected to IFOC condition (highlighted in Table \ref{tab:comp_diff_pertrb}) with the obtained ICOF results.
We observe that the variation of $\lm_{cr}$ with plate thickness ($\bar{h}$) and radius ratio ($B/A$) for both the boundary conditions is quite similar.
We also observe that for the plates subjected to ICOF boundary condition, the bifurcation solution associated with $R \text{-} \Theta \text{-} Z$ perturbation have minimum energy and is preferable over other type of perturbations.

\begin{figure}
\centering
\includegraphics[width=0.75\linewidth]{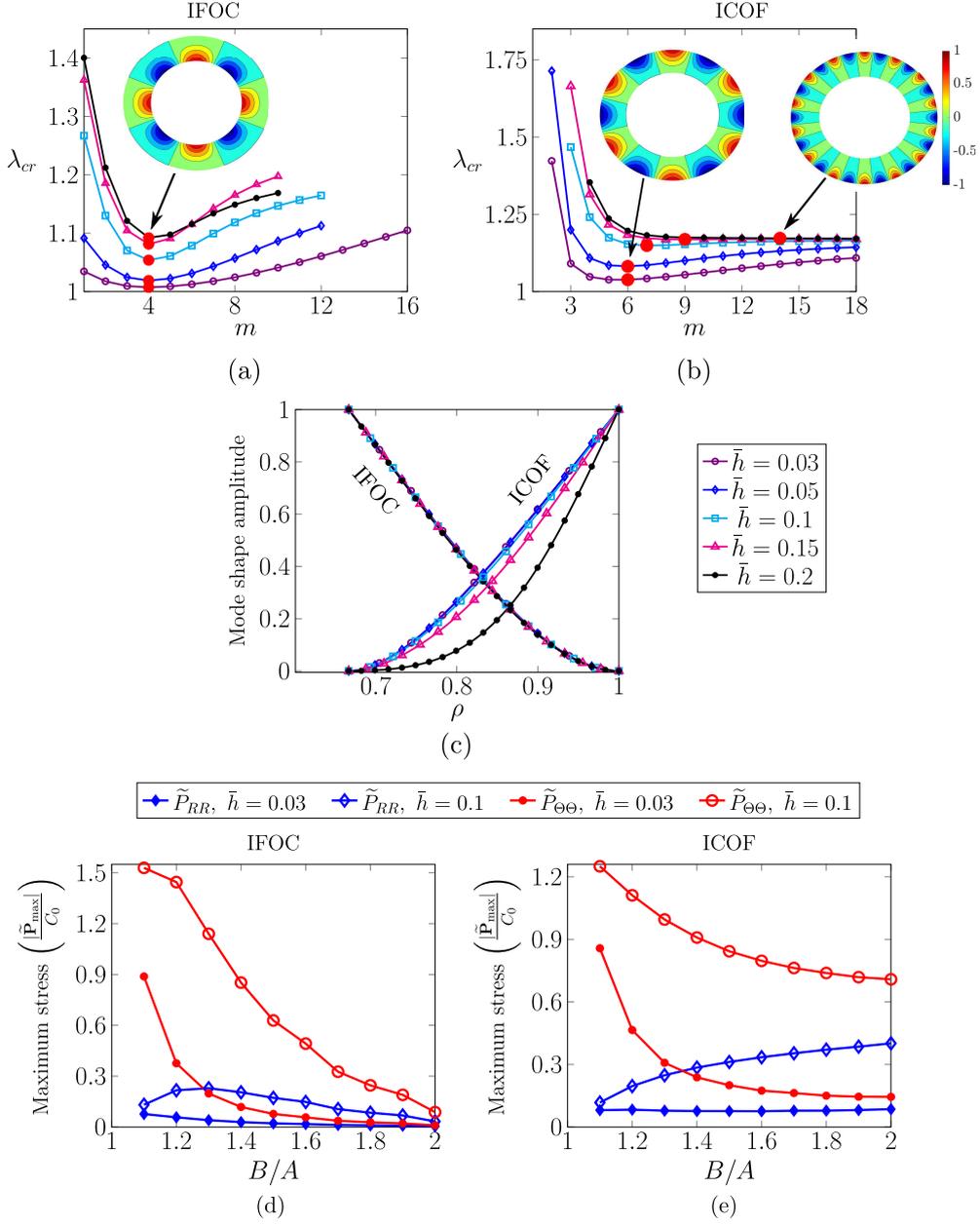}
\caption{ (a) and (b) Dependence of the critical growth factor $\lambda_{cr}$ on the wavenumber $m$ for an annular plate with aspect ratio $B/A = 1.5$ subjected to IFOC ($\Theta \text{-} Z$ perturbation) and ICOF ($R \text{-} \Theta \text{-} Z $ perturbation) boundary conditions, respectively. 
For each plate thickness value $\bar{h}$,  the red marks highlights the critical value of wavenumber ($m_{cr}$) corresponding to the lowest value of $\lm_{cr}$ on the solution curve. 
Normalised 2-D displacement contours at bifurcation for the IFOC boundary condition is plotted at $\bar{h} = 0.2,~ m_{cr} = 4,~ \lm_{cr} = 1.0922$. Similar contours for the ICOF boundary condition are plotted at $m_{cr} = 6, ~ \lm_{cr} = 1.0816$ and $m_{cr} = 14,~ \lm_{cr} = 1.1727$ for $\bar{h} = 0.05$ and $\bar{h} = 0.2$, respectively.
(c) Variation of the transverse amplitude of the mode shapes with the radius of the plate corresponding to the highlighted marks 
in (a) and (b).
For thick plate subjected to ICOF condition ($\bar{h} = 0.2, ~ m_{cr} = 14$), the radial extent of wrinkles is small suggesting that the wrinkles are more localised towards the outer unconstrained boundary of the plate. 
(d) and (e) Variation of normalised maximum stress components ($|\widetilde{\mbf{P}}_{\text{max}}|/C_0$) with plate aspect ratio and thickness for IFOC and ICOF case, respectively. 
The blue and red curve represents the radial ($ \widetilde{{P}}_{R R}$) and hoop ($ \widetilde{{P}}_{\Theta \Theta}$) stress distribution. The filled and unfilled markers represents the maximum stress at $\bar{h} = 0.03$ and $\bar{h} = 0.1$, respectively.} \label{fig:L_vs_m_IFOC_&_ICOF}
\end{figure}

Unlike the IFOC case, the values of $m_{cr}$ and $\lm_{cr}$ increase with $\bar{h}$ for all values of $B/A$. 
This can be visualised by plotting the variation of $\lm_{cr}$ with $m$ for various values of $\bar{h}$ in Figure \ref{fig:L_vs_m_IFOC_&_ICOF}a and Figure \ref{fig:L_vs_m_IFOC_&_ICOF}b. 
The critical wavenumber (corresponding to the minima of these curves) is constant with $\bar{h}$ for the annular plate ($B/A = 1.5, 2$) subjected to IFOC boundary condition as discussed in previous section. 
However, the $m_{cr}$ increases with the $\bar{h}$ for fixed $B/A$ using ICOF boundary condition. 
This suggest that a preferred bifurcation solution (satisfying the ICOF boundary conditions with $R\text{-}\Theta\text{-}Z$ perturbation) consist of more wrinkles in hoop direction for thicker plates at high value of $\lm_{cr}$. 
The contour plots with normalised displacement demonstrate the number of wrinkles in the circumferential direction at the bifurcation point. 
For the plate with $B/A = 1.5$ and $\bar{h} = 0.2$ subjected to IFOC boundary condition, the  critical wavenumber ($m_{cr}$) is 4 and does not change with the thickness. 
Whereas, for the plate of $B/A = 1.5$, $\bar{h} = 0.05$ subjected to ICOF boundary condition, the $m_{cr}$ is 6 and for the plate with $\bar{h} = 0.2$, the $m_{cr}$ is 14 which shows the increase in number of wrinkles with $\bar{h}$.
This is due to the expansion of the unconstrained outer boundary $b>B$ (considering ICOF) with the increase of thickness under growth stretch ($\lm>1$). The deformed configuration results in a larger circumference which can accommodate more number of wrinkles in the buckled configuration. 

Figure \ref{fig:L_vs_m_IFOC_&_ICOF}c shows the variation of transverse amplitude of the mode shape along the radius of the plate for different thickness at same radius ratio $B/A = 1.5$. In the case of IFOC, the mode shape variation remains the same for all plate thickness considered. In the case of ICOF, for thin plates ($m_{cr} = 6$), the radial extent of wrinkles is considerable i.e., the amplitude variation along the radius is more, whereas for thick plates the amplitude variation is localised at the unconstrained outer edge (as shown in Figure \ref{fig:L_vs_m_IFOC_&_ICOF}b). Thus, in the case of ICOF, the boundary layer effects are observed to be significant for thick plates when compared to thin plates.

Next, we investigate the effect of compressive stress on the wrinkle formation along the boundaries of annular plate. Figure \ref{fig:L_vs_m_IFOC_&_ICOF}d and Figure \ref{fig:L_vs_m_IFOC_&_ICOF}e shows the variation of normalised maximum stress ($\widetilde{\mbf{P}}_{\text{max}}/C_0$) with the plate radius ratio at different plate thickness values ($\bar{h} = 0.03,~ 0.1$) for IFOC and ICOF boundary conditions, respectively. The blue and red curves represents the dimensionless radial stress ($\widetilde{{P}}_{\text{rad}}/C_0 = \widetilde{P}_{RR}$) and circumferential stress ($\widetilde{{P}}_{\text{hoop}}/C_0 = \widetilde{{P}}_{\Theta \Theta}$), respectively. 
The variation of maximum stress with aspect ratio shows the similar behaviour for IFOC and ICOF case however, we observe that for IFOC condition, both the $\widetilde{P}_{R R}$ and $\widetilde{P}_{\Theta \Theta}$ are compressive which promotes the bifurcation where as for ICOF boundary condition, $\widetilde{P}_{\Theta \Theta}$ is compressive and $\widetilde{P}_{R R}$ is tensile that delays the bifurcation (Mathematical expressions for stresses are detailed in the supplementary document). Here, we plot the maximum stress value to show the dependence of stresses on the geometry of plate. Considering ICOF condition, for a fixed value of $B/A$, the value of critical growth stretch increases with the increase of thickness (due to increased bending stiffness), as shown in Table \ref{tab: comp_two_bound_at_critical_modes} which yields high compressive stresses as shown in Figure \ref{fig:L_vs_m_IFOC_&_ICOF}d and Figure \ref{fig:L_vs_m_IFOC_&_ICOF}e which results in more number of wrinkles. 
Whereas, the converse behaviour of stress is observed with respect to increase in the radius ratio of the annular plate for a fixed value of plate thickness that is the critical stretch value and compressive stresses decrease with the increase in radius ratio suggesting less wrinkles. For plates with low $B/A$ value, both the compressive stress and boundary layer effects govern the wrinkle formation. For plates with high $B/A$ value, only the compressive stress governs the wrinkle formation. Thus the combined effect of compressive stresses and boundary layer effects govern the number of wrinkles and their localization along the boundaries of the thick plate.
Also, the results show that the magnitude of $\lm_{cr}$ and $m_{cr}$ is higher for the bifurcation solution associated with ICOF case when compared to IFOC. 
For the same geometric parameters $B/A = 1.5$ and $\bar{h} = 0.1$, the value of $\lm_{cr}$ associated with IFOC condition is $1.0551$ at $m_{cr} = 4$ and ICOF condition is $\lm_{cr} = 1.1493$ at $m_{cr} = 7$.


\begin{table}
\centering
\caption{Dependence of the critical value of growth factor $(\lm_{cr})$ on plate thickness ($\bar{h}$), radius ratio ($B/A$), and critical wavenumber ($m_{cr}$). The lowest value of $\lm_{cr}$ is obtained for the annular plate with aspect ratio $B/A = 1.1,~ 1.5,~ 2$ at various plate thickness $\bar{h}$ subjected to IFOC and ICOF boundary conditions. The preferred bifurcation solution for ICOF boundary condition is obtained using $R \text{-} \Theta \text{-} Z$ perturbation.
For IFOC boundary condition, $R \text{-} \Theta \text{-} Z$  and $\Theta \text{-} Z$ perturbation are preferred for $B/A = 1.1$ and $B/A = 1.5,~2$, respectively.} \label{tab: comp_two_bound_at_critical_modes}
\begin{tabular}{p{1cm}| p{5cm}| p{5cm}| p{5cm}}
\hline
\hline
 $\bar{h}$  & 
 \begin{tabular}{c}
 $B/A = 1.1$ 
 \\
 \begin{tabular}{p{2cm} p{2cm}}
 \begin{tabular}{p{2cm}}
$\lm_{cr}$\\
\small{(IFOC)}
\end{tabular}  
& 
\begin{tabular}{p{2.5cm}}
$\lm_{cr}$\\
\small{(ICOF)}
\end{tabular}
\end{tabular} 
\end{tabular}
&
 \begin{tabular}{c}
  $B/A = 1.5$ 
 \\
 \begin{tabular}{p{2cm} p{2cm}}
 \begin{tabular}{p{2.5cm}}
$\lm_{cr}$\\
\small{(IFOC)}
\end{tabular}  
& 
\begin{tabular}{p{2.5cm}}
$\lm_{cr}$\\
\small{(ICOF)}
\end{tabular}
\end{tabular} 
\end{tabular}
&
 \begin{tabular}{c}
  $B/A = 2$ 
 \\
 \begin{tabular}{p{2cm} p{2cm}}
 \begin{tabular}{p{2.5cm}}
$\lm_{cr}$\\
\small{(IFOC)}
\end{tabular}  
& 
\begin{tabular}{p{2.5cm}}
$\lm_{cr}$\\
\small{(ICOF)}
\end{tabular}
\end{tabular} 
\end{tabular}
\\
\hline
0.03 & \begin{tabular}{p{2cm} p{2cm}}
\begin{tabular}{p{3cm}}
1.1127 \\
($m_{cr} = 20$)\\
($R \text{-} \Theta \text{-} Z$)
\end{tabular}
& 
\begin{tabular}{p{3cm}}
1.1327\\
($m_{cr} = 23$)\\
($R \text{-} \Theta \text{-} Z$)
\end{tabular}
\end{tabular}
&
\begin{tabular}{p{2cm} p{2cm}}
\begin{tabular}{p{3cm}}
1.0074 \\
 ($m_{cr} = 4$)\\
($\Theta \text{-} Z$)
\end{tabular}
 & 
\begin{tabular}{p{3cm}}
 1.0386\\
  ($m_{cr}= 6$)\\
  ($R \text{-} \Theta \text{-} Z$)
\end{tabular} 
\end{tabular}
&
\begin{tabular}{p{2cm} p{2cm}}
\begin{tabular}{p{3cm}}
1.0007\\
($m_{cr} = 2$)\\
  ($\Theta \text{-} Z$)
\end{tabular}
&
\begin{tabular}{p{3cm}}
1.0306\\
($m_{cr} = 4$)\\
  ($R \text{-} \Theta \text{-} Z$)
\end{tabular}
\end{tabular}
\\
\hline
0.05 & 
\begin{tabular}{p{2cm} p{2cm}}
\begin{tabular}{p{3cm}}
1.1547\\
($m_{cr} = 21$)\\
($R \text{-} \Theta \text{-} Z$)
\end{tabular}
&
\begin{tabular}{p{3cm}}
1.1680\\
($m_{cr} = 26$)\\
($R \text{-} \Theta \text{-} Z$)
\end{tabular}
\end{tabular}
&
\begin{tabular}{p{2cm} p{2cm}}
\begin{tabular}{p{3cm}}
1.0191 \\
($m_{cr} = 4$)\\
($\Theta \text{-} Z$)
\end{tabular}
 & 
\begin{tabular}{p{3cm}}
 1.0816\\
  ($m_{cr} = 6$)\\
  ($R \text{-} \Theta \text{-} Z$)
\end{tabular} 
\end{tabular}
&
\begin{tabular}{p{2cm} p{2cm}}
\begin{tabular}{p{3cm}}
1.0019 \\
($m_{cr} = 2$)\\
  ($\Theta \text{-} Z$)
\end{tabular}
&
\begin{tabular}{p{3cm}}
1.0681\\
($m_{cr} = 4$)\\
  ($R \text{-} \Theta \text{-} Z$)
\end{tabular}
\end{tabular}
\\
\hline
0.1 & 
\begin{tabular}{p{2cm} p{2cm}}
\begin{tabular}{p{3cm}}
1.1793 \\
 ($m_{cr} = 22$)\\
($R \text{-} \Theta \text{-} Z$)
\end{tabular}
& 
\begin{tabular}{p{3cm}}
1.1850\\
 ($m_{cr} = 30$)\\
($R \text{-} \Theta \text{-} Z$)
\end{tabular}
\end{tabular}
&
\begin{tabular}{p{2cm} p{2cm}}
\begin{tabular}{p{3cm}}
 1.0551 \\
 ($m_{cr}= 4$)\\
($\Theta \text{-} Z$)
\end{tabular}
 & 
\begin{tabular}{p{3cm}}
 1.1493\\
  ($m_{cr} = 7$)\\
  ($R \text{-} \Theta \text{-} Z$)
\end{tabular} 
\end{tabular}
&
\begin{tabular}{p{2cm} p{2cm}}
\begin{tabular}{p{3cm}}
1.0063 \\
($m_{cr} = 2$)\\
  ($\Theta \text{-} Z$)
\end{tabular}
&
\begin{tabular}{p{3cm}}
1.1387\\
($m_{cr}= 5$)\\
  ($R \text{-} \Theta \text{-} Z$)
\end{tabular}
\end{tabular}
\\
\hline
0.15 & 
\begin{tabular}{p{2cm} p{2cm}}
\begin{tabular}{p{3cm}}
1.1843\\
 ($m_{cr} = 23$)\\
($R \text{-} \Theta \text{-} Z$)
\end{tabular}
 & 
\begin{tabular}{p{3cm}}
1.1879 \\
 ($m_{cr} = 32$)\\
($R \text{-} \Theta \text{-} Z$)
\end{tabular} 
\end{tabular}
&
\begin{tabular}{p{2cm} p{2cm}}
\begin{tabular}{p{3cm}}
 1.0814\\
  ($m_{cr} = 4$)\\
($\Theta \text{-} Z$)
\end{tabular}
  & 
  \begin{tabular}{p{3cm}}
   1.1690 \\
    ($m_{cr} = 9$)\\
  ($R \text{-} \Theta \text{-} Z$)
  \end{tabular}
\end{tabular}
&
\begin{tabular}{p{2cm} p{2cm}}
\begin{tabular}{p{3cm}}
1.0113 \\
($m_{cr} = 2$)\\
  ($\Theta \text{-} Z$)
\end{tabular}
&
\begin{tabular}{p{3cm}}
1.1624\\
($m_{cr}= 6$)\\
  ($R \text{-} \Theta \text{-} Z$)
\end{tabular}
\end{tabular}
\\
\hline
0.2 &
\begin{tabular}{p{2cm} p{2cm}}
\begin{tabular}{p{3cm}}
1.1860\\
 ($m_{cr} = 23$)\\
($R \text{-} \Theta \text{-} Z$)
\end{tabular}
 &   
\begin{tabular}{p{3cm}}
1.1888 \\
($m_{cr} = 48$)\\
($R \text{-} \Theta \text{-} Z$)
\end{tabular}   
\end{tabular}
&
\begin{tabular}{p{2cm} p{2cm}}
\begin{tabular}{p{3cm}}
 1.0962\\
  ($m_{cr} = 4$)\\
($\Theta \text{-} Z$)
\end{tabular}
  &   
\begin{tabular}{p{3cm}}
1.1727\\
($m_{cr}= 14$)\\
  ($R \text{-} \Theta \text{-} Z$)
\end{tabular}  
\end{tabular}
&
\begin{tabular}{p{2cm} p{2cm}}
\begin{tabular}{p{3cm}}
1.0155\\
($m_{cr} = 2$)\\
  ($\Theta \text{-} Z$)
\end{tabular}
&
\begin{tabular}{p{3cm}}
1.1705\\
($m_{cr} = 6$)\\
  ($R \text{-} \Theta \text{-} Z$)
\end{tabular}
\end{tabular}
\\
\hline
\end{tabular}
\end{table}

\subsubsection{Comparison with computational results for growing annulus}
We have discussed in Section \ref{sec:comp_circular_ring} that current plate theory estimates pre-buckling results for circular ring and annular shell. 
Now, to test the accuracy of numerical framework, we compare the bifurcation solution of the annular plate using current theory with the computational results for a nearly incompressible (Poisson ratio = 0.495) neo-Hookean growing annulus provided by \cite{groh2022morphoelastic} using the finite element method. 
He used a  seven-parameter quadrilateral shell element to analyse the growth-induced instability in thin growing shell by implementing the numerical continuation algorithm.
In his work, the isotropic planar growth tensor is given as $\mbf{G} = \text{diag}(1+\lm_g, 1+ \lm_g, 1)$. 
The prescribed boundary condition were pinned inner edge and free outer edge.
He reported the critical value of growth function and wavenumber for growing annulus as $\lm_{gcr} = 0.137 \times 10^{-5}$ and $m_{cr} = 3$, respectively. 
For the same parameters, radius ratio ($B/A = 2$), shell thickness ($2 h = 0.001 \rightarrow 2\bar{h} = 0.0005$), and the boundary conditions which are $U(A^*) = V(A^*) = W(A^*) = W''(A^*) = 0$ (corresponding to simply supported inner boundary) and $U'(1) = V'(1) = W''(1) = W'''(1) =0 $ (corresponding to free outer boundary), the current incompressible plate theory yields the critical wavenumber $m_{cr} = 3$ with $\lm_{gcr} = 0.152 \times 10^{-5}$.
The two results are relatively close to each other with a small deviation arising due to the non satisfaction of incompressibility constraint in \cite{groh2022morphoelastic}'s formulation.

\section{Conclusion} \label{conclusion}
In this work, we have investigated the wrinkling phenomena in growing hyperelastic annular plates using a finite strain asymptotic plate theory. 
A 3-D plate equilibrium system is reduced to 2-D plate governing system by adopting series expansion along the thickness direction.
A homogeneous isotropic growth function is considered as a control parameter in inducing the  circumferential instability in an incompressible neo-Hookean annular plate. 
To validate the 2-D plate framework, we compared the numerical pre-buckling solution for a very thin annular plate with the analytical pre-buckling solution of circular ring. Both the analytical and numerical results are in good agreement.
We carried out linear bifurcation analysis with asymmetric  (i.e., along $R \text{-} \Theta \text{-} Z$ and $\Theta \text{-} Z$ direction) as well as axisymmetric perturbations (i.e., along $R \text{-} Z$ direction) for two cases of boundary conditions (IFOC and ICOF). The numerical solution of resulting system of ODEs in each case is solved using the compound matrix method. The critical value of growth factor ($\lm_{cr}$) and the associated wavenumber ($m_{cr}$) is evaluated for each type of perturbation and boundary conditions.
We observe that the bifurcation solution corresponding to asymmetric perturbation is preferred for both boundary conditions as it has a lower value of $\lm_{cr}$ when compared to the axisymmetric perturbation. 
The bifurcation solutions associated with IFOC and ICOF boundary conditions exhibit a similar behaviour for variation of $\lm_{cr}$ (at $m_{cr}$) with $B/A$ and $\bar{h}$. However, the magnitude of $\lm_{cr}$ is lower for the case of IFOC when compared to ICOF  due to the presence of compressive radial and circumferential stresses which promotes wrinkling.
In addition, for ICOF case, we find that for a fixed value of $B/A$, the deformed radius and $\lm_{cr}$ increases with plate thickness resulting in higher compressive stress in circumferential direction which further results in more number of localised wrinkles along the outer circumference of the thicker plate. 
To test the accuracy of obtained results for annular plate, we compare the bifurcation solution obtained for incompressible annular plate with the existing bifurcation solution for slightly incompressible shell obtained using finite element approach. Both the bifurcation results are close to each other showing the consistency of current plate theory.

Furthermore, we have restricted our study to determine the critical value of growth factor responsible for the onset of wrinkling, however a post-bifurcation analysis may provide insights on the evolution of wrinkle deformation with growth. This is currently being investigated and our findings will be reported in a suitable forum at a later stage.

\section*{Acknowledgements}
Prashant Saxena acknowledges the financial support of the EPSRC grant no. EP/V030833/1.

\addcontentsline{toc}{section}{References}
\bibliographystyle{author-year-prashant}
\bibliography{growing_annular_plate} 

\begin{appendix}

\section{Appendix: Expression for Piola stress and unknown variables} \label{app:series_exp_for_tensor}
The series expansion of deformation gradient ($\mbf{F}$), elastic deformation ($\mbf{A}$), inverse transpose of Growth tensor ($\mbf{G}^{-T}$) and Piola Kirchhoff ($\mbf{P}$) tensor are given by 
\begin{equation}
\begin{aligned}
& \mbf{F} = \sum_{n = 0}^{2} \frac{Z^n}{n!} \mbf{F}^{(n)}(\zeta) + O(Z^3), \quad  & \mbf{A} = \sum_{n = 0}^{2} \frac{Z^n}{n!} \mbf{A}^{(n)}(\zeta) + O(Z^3),  \\
& \mbf{G}^{-T} = \sum_{n = 0}^{2} \frac{Z^n}{n!} {\bar{\mbf{G}}}^{n}(\zeta) + O(Z^3), \quad & \mbf{P} = \sum_{n = 0}^{2} \frac{Z^n}{n!} \mbf{P}^{(n)}(\zeta) + O(Z^3).
\end{aligned} \label{app:series_exp}
\end{equation}
For an incompressible neo-Hookean material elastic strain energy function is $\phi_0(\mathbf{A})=C_0[\text{tr}(\mathbf{A}^T\mathbf{A})-3]$ and the associated Piola Kirchhoff stress is given as,
$\mathbf{P}=J_G \left[2C_0[\mathbf{A}]-p \mathbf{A}^{-T}\right]\mathbf{G}^{-T}$. 
Then, the first term in right side of the expression for $\mbf{P}$ in \eqref{app:series_exp} is obtained as
\begin{align}
\mathbf{P}^{(0)} = J_G\left[2C_0\mathbf{A}^{(0)}-p{\mathbf{A}^{(0)}}^{-T}\right]\bar{\mathbf{G}}^{(0)}. \label{app:exp_for_P0}
\end{align}
By using bottom traction condition, $\mathbf{P}^{(0)} \mathbf{k}= \mbf{0}$ and substituting the expression for $\mathbf{A}^{(0)}$ (see \eqref{eq:exp_elastic_def}) in \eqref{app:exp_for_P0} we obtain
\begin{align}
2C_0\nabla \mathbf{x}^{(0)}\bar{\mathbf{G}}^{{(0)}^{T}} \widehat{\mathbf{G}}^{(0)}\mathbf{k}+2C_0 J_G \left|\bar{\mathbf{G}}^{(0)}\mathbf{k}\right|^2\mathbf{x}^{(1)}-p^{(0)} {\mbf{F^{(0)}}}^*\mathbf{k} = \mathbf{0}, \label{eq:trac_exp}
\end{align}
where $J^{(0)}=\left. J_G \right|_{Z=0}$, $\widehat{\mbf{G}}^{(0)} = J^{(0)} \bar{\mbf{G}}^{(0)}$, and ${\mbf{F^{(0)}}}^* = $  $\text{Cofac}(\mathbf{F}^{(0)})$. In this work, we use $\nabla \mathbf{x}^{{(0)}^*}$ in place of $\mbf{F^{(0)}}^* \mathbf{k}$ which is given as
\begin{align} \label{eq:exp_cofac_k}
\nabla \mathbf{x}^{{(0)}^*} = \displaystyle  \frac{r^{(0)}}{R}\left[\frac{\partial \theta^{(0)}}{\partial R}\displaystyle\frac{\partial z^{(0)}}{\partial \Theta}- \displaystyle\frac{\partial \theta^{(0)}}{\partial \Theta}\displaystyle\frac{\partial z^{(0)}}{\partial R}\right] \mathbf{e}_1+\displaystyle \frac{1}{R}\left[\displaystyle\frac{\partial r^{(0)}}{\partial \Theta}\displaystyle\frac{\partial z^{(0)}}{\partial R}-\displaystyle\frac{\partial r^{(0)}}{\partial R}\displaystyle\frac{\partial z^{(0)}}{\partial \Theta}\right]\mathbf{e}_2 \nonumber \\
+ \displaystyle \frac{r^{(0)}}{R} \left[\displaystyle\frac{\partial r^{(0)}}{\partial R}\displaystyle\frac{\partial \theta^{(0)}}{\partial \Theta}-\displaystyle\frac{\partial \theta^{(0)}}{\partial R}\displaystyle\frac{\partial r^{(0)}}{\partial \Theta}\right]\mathbf{k}. 
\end{align}
Using incompressibility constraint det$(\mbf{A}) = 1$, we obtain $ \displaystyle \text{det}(\mathbf{F}^{(0)})=\text{det}(\bar{\mathbf{G}}^{{(0)}^{-T}})$ which result in
\begin{align}
\mathbf{x}^{(1)} \cdot \nabla \mathbf{x}^{{(0)}^*}=\det\left(\bar{\mathbf{G}}^{{(0)}^{-T}}\right), \label{eq:det_definition}
\end{align}
where $\displaystyle \text{det}({\mathbf{F}}^{(0)})=\bigg[r^{(1)}\mathbf{e}_1 + r^{(0)} \theta^{(1)} \mathbf{e}_2 + z^{(1)} \mathbf{e}_3 \bigg] \cdot \mbf{F^{(0)}}^*\mathbf{k}= \mathbf{x}^{(1)} \cdot \nabla \mathbf{x}^{{(0)}^*}$.
Using \eqref{eq:trac_exp} we obtain the explicit expression for $\mathbf{x}^{(1)}$ 
\begin{align}
\mathbf{x}^{(1)}&=\frac{-2C_0\nabla \mathbf{x}^{(0)}\bar{\mathbf{G}}^{{(0)}^{T}} \widehat{\mathbf{G}}^{(0)}\mathbf{k}+p^{(0)}\nabla \mathbf{x}^{{(0)}^*}}{2C_0 J_G \left|\bar{\mathbf{G}}^{(0)}\mathbf{k}\right|^2}.\label{x1_exp}
\end{align}
To obtain the explicit expression for $p^{(0)}$ we substitute \eqref{x1_exp} into \eqref{eq:det_definition} which yields 
\begin{align}
p^{(0)}=\frac{2C_0 J_G \left|\bar{\mathbf{G}}^{(0)}\mathbf{k}\right|^2}{\text{det}\bar{\mathbf{G}}^{{(0)}^{T}}\left|\nabla \mathbf{x}^{{(0)}^*}\right|^{2}}+\left[ 2 C_0\nabla \mathbf{x}^{(0)}\bar{\mathbf{G}}^{{(0)}^{T}} \widehat{\mathbf{G}}^{(0)}\mathbf{k}\right] \cdot \frac{\nabla \mathbf{x}^{{(0)}^*}}{{\left|\nabla \mathbf{x}^{{(0)}^*}\right|^{2}}}.\label{eq:p0_exp}
\end{align}
Using Eq. \eqref{eq:p0_exp}, we obtain the expression for $p^{(0)}$ which is
\begin{align}
p^{(0)}=\frac{2C_0 \lambda^4}{\left|\nabla {\mathbf{x}^{(0)}}^{*}\right|^2},
\end{align}
where $\nabla{\mbf{x}^{(0)}}^{*}= \Delta {x}_{11} \mathbf{e}_1 + \Delta {x}_{22} \mathbf{e}_2 + \Delta {x}_{33} \mathbf{k}$ is given by \eqref{eq:exp_cofac_k}.
On substituting $p^{(0)}$ in \eqref{x1_exp} we obtain explicit expressions for $r^{(1)}$, $\theta^{(1)}$, and $z^{(1)}$ as
\begin{align}
r^{(1)}= \displaystyle \frac{p^{(0)} \Delta x_{11}}{2 C_0 \lambda^2}, \quad
\theta^{(1)}=\frac{p^{(0)} \Delta x_{22}}{2 C_0 \lambda^2 r^{(0)}}, \quad \text{and} \quad 
z^{(1)}=\frac{p^{(0)} \Delta x_{33}}{2 C_0 \lambda^2}.
\end{align} 

\end{appendix}

\end{document}